\documentclass[preprint,12pt]{elsarticle}

\usepackage{amssymb}
\usepackage{epsfig}
\usepackage{amsthm}
\usepackage{color}
\usepackage{amsfonts}
\usepackage{graphicx}
\usepackage{amsmath}
\usepackage[all]{xy}

\journal{ArXiv}

\begin{document}

\begin{frontmatter}

\title{Electron-positron planar orbits in a constant magnetic field}

\author{M.A. Gonzalez Leon$^1$, J. Mateos Guilarte$^2$ and M. de la Torre Mayado$^2$}

\address{$^1$Departamento de Matem\'atica Aplicada and IUFFyM, University of Salamanca, Spain\\ $^2$Departamento de F\'{\i}sica Fundamental and IUFFyM, University of Salamanca, Spain}

\begin{abstract}
The different types of orbits in the classical problem of two particles with equal masses and opposite charges on a plane under the influence of a constant orthogonal magnetic field are classified. The equations of the system are reduced to the problem of a Coulomb center plus a harmonic oscillator. The associated bifurcation diagram is fully explained. Using this information the dynamics of the two particles is described.
\end{abstract}




\end{frontmatter}

\section{Introduction}

The problem of two charged particles moving on a plane in presence of a constant perpendicular magnetic field has been profusely analyzed in both classical and quantum context, see   \cite{Avron}-\cite{Escobar}. This dynamical system is related to several interesting physical problems as the Quantum Hall Effect \cite{Curilef} or the diamagnetic Kepler problem, see \cite{Pinheiro} and references therein. In the particular case of equal masses and opposite charges, for instance electron and positron, the classical equations can be reduced to a completely integrable system \cite{Gadella} describing the motion of a particle in presence of a Coulomb center and a harmonic oscillator. In this work we classify and describe the different types of orbits for this classical problem. In particular we focus on the structure of the bifurcation diagram associated to the reduced system.

The general system with arbitrary masses and charges has been completely analyzed in \cite{Pinheiro} for the planar case and \cite{Pinheiro2} for the three-dimensional one. Different combinations of masses and charges give rise to distinct behaviour with respect to the integrability and solvability properties for the corresponding equations. A detailed study of special solutions has been performed in \cite{Turbiner}, and also in \cite{Escobar} for the more general case of three charged particles in the plane. The problem is developed from the quantum point of view in \cite{Turbiner2,Turbiner3,Gadella2}.

The planar system presents four independent first integrals: energy, total angular momentum and the two components of the magnetic translations momentum corresponding to the center of mass for the two particles. However these constants of motion neither are in involution nor close a Lie algebra structure. For the particular case of equal masses and charges, the dynamics is uncoupled in the center of mass reference frame, and thus the system equations split in two independent integrable systems \cite{Curilef,Pinheiro}. The electrically neutral case with equal masses has an interesting property: even though the equations of the system are coupled, the components of the total magnetic momentum are in involution. This result permits a reduction process from the original system to a two-dimensional problem that is Hamilton-Jacobi separable in elliptic coordinates \cite{Gadella}, and thus completely integrable.

We shall analyze in full detail the bifurcation diagram for the reduced problem and the corresponding regions for allowable motion. The final product of this analysis will be a complete classification for the different types of orbits in both the reduced and the original physical problem.

The structure of the paper is as follows: in Section 2 the system is described and its symmetries are presented. Section 3 is devoted to develop the reduction process and the description of the separability for the relative equations of motion in Euler elliptic coordinates, i.e. a Liouville system of type I \cite{Perelomov}. In Section 4 the bifurcation diagram is constructed and its structure explained, giving rise to the classification of the different types of orbits for the reduced system. Finally, in Section 5, the dynamics of the original problem is described using the previous analysis of the bifurcation diagram. A numerical approach is used to unveil the behaviour of the electron-positron orbits and a complete gallery of graphical representations is offered in an Appendix.  The influence of physical parameters and initial conditions is also analyzed.


\section{Electron-positron dynamics in a constant magnetic field}

Let us consider the Hamiltonian system $(M, H, \omega)$ describing two charged particles moving in a plane, with the same mass $m_1=m_2=m$ and opposite
charges $e_1=-e_2=e \ , \ e>0$, under the influence of a constant magnetic field ${\bf B} = - B {\bf k}$. $M$ denotes the phase space, i.e. $M = T^*(\mathbb{R}^2 \times \mathbb{R}^2)$, and $\omega$ stands for the canonical symplectic form in $M$:
\[
\omega = d x_1 \wedge d p_{x_1} + d y_1 \wedge d p_{y_1} + d x_2 \wedge d p_{x_2} + d y_2 \wedge d p_{y_2}  \  ,
\]
$(x_1,y_1)$ and $(x_2,y_2)$ being the respective positions of the two particles, and $(p_{x_1},p_{y_1}), (p_{x_2},p_{y_2})$ the associated canonical momenta. The Hamiltonian function is written as
\begin{equation}
H = H_1 + H_2 + H_{12}\label{hsum}
\end{equation}
where:
\begin{eqnarray}
H_1 &=& \frac{1}{2m} \left[ \left( p_{x_1} - \frac{eB}{2c} y_1 \right)^2 + \left( p_{y_1} + \frac{eB}{2c} x_1 \right)^2 \right] \label{h1} \\
H_2 &=& \frac{1}{2m} \left[ \left( p_{x_2} + \frac{eB}{2c} y_2 \right)^2 + \left( p_{y_2} - \frac{eB}{2c} x_2 \right)^2  \right] \label{h2} \\
H_{12} &=& - \frac{e^2}{\sqrt{(x_1-x_2)^2+(y_1-y_2)^2}} \label{h12}
\end{eqnarray}
and the vector potential has been chosen in the symmetric gauge, ${\bf A}(x_i,y_i) = \left( \frac{B}{2} y_i, -\frac{B}{2} x_i \right)$, $i = 1, 2$. Non-rationalized Gauss units so that $[e^2]=M L^3 T^{-2}$ and $[eB]= M L T^{-2}$ have been used in (\ref{h1},\ref{h2},\ref{h12}).

Hamiltonians (\ref{h1}) and (\ref{h2}) describe two classical Landau problems in the symmetric gauge, each of them exhibits as fundamental symmetries magnetic translations and  invariance under rotations \cite{Florek,Wal,Zak}. The interaction term (\ref{h12}) breaks these symmetries, but the Hamiltonian (\ref{hsum}) is invariant under total magnetic translations and rotations of the complete physical system \cite{Avron,Pinheiro,Turbiner}. Thus the conserved quantities for the system (\ref{hsum}) are the total magnetic translations momenta ($K_X$, $K_Y$) and the total angular momentum $L_Z$, defined by:
\begin{eqnarray}
K_X &=& k_{x_1} + k_{x_2} = \left( p_{x_1} + \frac{eB}{2c} y_1\right) +\left(  p_{x_2} - \frac{eB}{2c} y_2 \right) \label{tmTX} \\
K_Y &=& k_{y_1} + k_{y_2} = \left( p_{y_1} - \frac{eB}{2c} x_1 \right) +\left( p_{y_2} + \frac{eB}{2c} x_2 \right) \label{tmTY}\\
L_Z &=& L_{z_1} + L_{z_2} =\left( x_1 p_{y_1} - y_1 p_{x_1} \right) + \left( x_2 p_{y_2} - x_2 p_{y_2}  \right)\label{maT}\nonumber
\end{eqnarray}

Here $k_{x_i}$, $k_{y_i}$, $i = 1, 2$, generate the magnetic translations for the two particles respectively, and $L_{z_i}$ are the
angular momenta in the orthogonal direction to the plane, for each particle. Thus:
\[
\{K_X, H\} = \{K_Y, H\}= \{L_Z, H\} =0
\]
where $\{ \cdot,\cdot \}$ is defined as the bracket canonically associated to $\omega$. It is remarkable that $L_Z$ is not in involution with $K_X$ and $K_Y$:
\[
\{K_X, L_Z\} = - K_Y \  ;\quad
\{K_Y, L_Z\} = \ K_X
\]
whereas: $\{K_X, K_Y\} = 0$ for this special case of electrically neutral system \cite{Avron,Pinheiro,Turbiner}.

Hamilton equations for the Hamiltonian (\ref{hsum}) give the expression of the kinematical momenta:
\begin{equation}
\begin{array}{ll}\displaystyle  m \dot{x}_1 = p_{x_1}- \frac{eB}{2c} y_1, \quad & \displaystyle m \dot{y}_1 = p_{y_1}+ \frac{eB}{2c} x_1  \\ & \\
\displaystyle m \dot{x}_2 = p_{x_2}+ \frac{eB}{2c} y_2, \quad &\displaystyle  m \dot{y}_2 = p_{y_2}- \frac{eB}{2c} x_2 \end{array}\nonumber
\end{equation}
and reproduce Newton's equations for the system:
\begin{equation}
\begin{array}{ll} \displaystyle m \ddot{x}_1 = - \frac{eB}{c} \dot{y}_1 - \frac{e^2}{r^3} (x_1-x_2),\quad  &
\displaystyle  m \ddot{y}_1 = \ \ \frac{eB}{c} \dot{x}_1 - \frac{e^2}{r^3} (y_1-y_2) \\ &\\
\displaystyle  m \ddot{x}_2 = \ \ \frac{eB}{c} \dot{y}_2 + \frac{e^2}{r^3} (x_1-x_2), \quad &
\displaystyle  m \ddot{y}_2 = - \frac{eB}{c} \dot{x}_2 + \frac{e^2}{r^3} (y_1-y_2)
\end{array} \label{Newtoneqs}
\end{equation}
where $r = \sqrt{(x_1-x_2)^2+(y_1-y_2)^2}$.

The integrability of this system has been analyzed by several authors, see for instances \cite{Curilef,Pinheiro,Gadella,Turbiner} and references therein. It is natural to consider in this kind of systems the center of mass reference frame and the associated relative particle. As we will see in next section, the motions of the center of mass and the relative particle are not independent. However invariance under global magnetic translations lead to a reduction process that allows to convert the equations (\ref{Newtoneqs}), for fixed values of the magnetic momenta, into the Newton equations for a particle in presence of a Coulomb center and a harmonic oscillator \cite{Curilef,Gadella}.


\section{Equivalence with a planar Liouville type I system}

\subsection{Reduction}
\label{reduction}

Rearranging Newton equations (\ref{Newtoneqs}), this ODE system is re-writen in the equivalent way:
\begin{eqnarray}
&& m (\ddot{x}_1 +\ddot{x}_2) =  \ \ \frac{eB}{c} (\dot{y}_2 - \dot{y}_1)\label{NLcm1} \\
&& m (\ddot{y}_1 +\ddot{y}_2) = - \frac{eB}{c} (\dot{x}_2 - \dot{x}_1) \label{NLcm2}\\
&& m (\ddot{x}_2 -\ddot{x}_1) =  \ \ \frac{eB}{c} (\dot{y}_1 + \dot{y}_2)-\frac{2e^2}{r^3} (x_2-x_1)\label{NLpr1} \\
&& m (\ddot{y}_2 -\ddot{y}_1) = - \frac{eB}{c} (\dot{x}_1 + \dot{x}_2) -\frac{2e^2}{r^3} (y_2-y_1) \label{NLpr2}
\end{eqnarray}
where it is clear that (\ref{NLcm1}) and (\ref{NLcm2}) are no more than the conservation equations for $K_X$ and $K_Y$, see (\ref{tmTX},\ref{tmTY}). In fact, introducing the center of mass coordinates for the system: $X=\frac{1}{2} (x_1+x_2)$ and $Y=\frac{1}{2} (y_1+y_2)$, and the relative coordinates: $x=x_2-x_1$ and $y=y_2-y_1$, (\ref{NLcm1}) and (\ref{NLcm2}) are:
\begin{eqnarray}
&& 2 m \ddot{X}  =  \ \ \frac{eB}{c}\dot{y}\quad \Longrightarrow \frac{d}{dt} K_X = 0 \label{CL1} \\
&& 2 m \ddot{Y} = - \frac{eB}{c} \dot{x} \quad  \Longrightarrow \frac{d}{dt} K_Y = 0 \label{CL2}
\end{eqnarray}
Integrating (\ref{CL1}) and (\ref{CL2}), the values of $K_X$ and $K_Y$ will be fixed: $K_X = K_1$, $K_Y=K_2$, $\ K_1, K_2 \in \mathbb{R}$. Substituting these values of the constants of motion in equations (\ref{NLpr1}) and (\ref{NLpr2}), we obtain:
\begin{eqnarray}
\mu \ddot{x} &=&  - \mu {\rm \omega}^2 (x-x_0) - \frac{e^2}{r^3} x \label{NLr1a} \\
\mu \ddot{y} &=& - \mu {\rm \omega}^2 (y+y_0) - \frac{e^2}{r^3}y  \label{NLr2b}
\end{eqnarray}
where $x_0=\frac{K_2}{m\omega}$, $y_0=\frac{K_1}{m\omega}$, $\mu = \frac{m}{2}$ and  $\omega=\frac{eB}{mc}$ is the cyclotron frequency.

Thus, for any given pair of values $K_1$ and $K_2$, the equations of system (\ref{hsum}) are reduced to a system of two ODE, (\ref{NLr1a}, \ref{NLr2b}), depending only on the relative coordinates $(x,y)$. Equations (\ref{NLr1a}, \ref{NLr2b}) can be easily re-interpreted as the motion equations of a particle on a plane, of mass $\mu$, with a potential energy:
\begin{equation}
U(x,y) = \frac{1}{2} \mu \omega^2 ((x-x_0)^2+(y+y_0)^2) -\frac{e^2}{\sqrt{x^2+y^2}} \nonumber \label{potr}
\end{equation}
i.e. the superposition of a harmonic oscillator centered in $(x_0,-y_0)$ and a Coulomb potential in the origin of coordinates \cite{Curilef,Gadella}. Moreover, Gadella \textit{et al.} have proved in \cite{Gadella} that this system is separable in elliptic coordinates, and thus completely integrable. Here, we will construct the complete separation process, in Euler elliptic coordinates, in order to obtain the bifurcation diagram for the reduced system. Its analysis will permit the complete classification of the different types of motions for the reduced problem, and,  consequently, for the original system of an electron and a positron in a constant magnetic field.

\begin{figure}
\begin{center}
  \includegraphics[height=6cm]{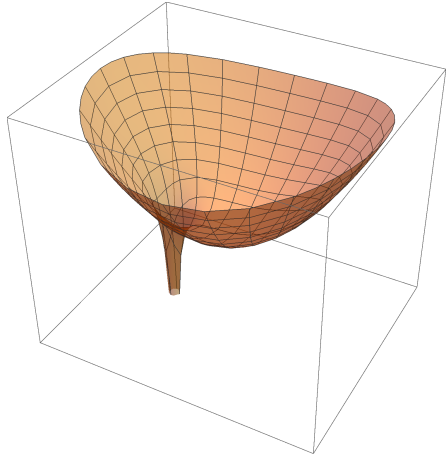}
\caption{Graphics of the reduced potential (\ref{redham}) for $\alpha=\frac{1}{3}$ and $a=1$.}
\label{potred}       
\end{center}
\end{figure}

It is convenient to use dimensionless variables:
\[
x \rightarrow l x \  , \quad y \rightarrow l y \  , \quad  p_x \rightarrow \mu \omega l  p_x \   , \quad  p_y \rightarrow \mu \omega  l p_y \   , \quad  t \rightarrow \omega \, t
\]
where the magnetic length has been chosen as $l=\frac{mc}{eB}$, i.e. the speed-one gyromagnetic radius. We also introduce an affine transformation:
\[
q_1 = \frac{x_0}{a} x-\frac{y_0}{a} y -a\  ,\quad q_2=\frac{y_0}{a} x+\frac{x_0}{a} y\  ,
\]
in such a way that the harmonic oscillator is now located at the origin, whereas the Coulomb center is displaced to the point $(-a,0)$, where $a=\sqrt{x_0^2+y_0^2}$. The transformed dimensionless Hamiltonian for the reduced system will read (see Fig. \ref{potred}):
\begin{equation}
{\cal H} = \frac{1}{2}(p_1^2 + p_2^2) +  \frac{1}{2} (q_1^2+q_2^2) - \frac{\alpha}{\sqrt{(q_1+a)^2 + q_2^2}} \label{redham}
\end{equation}
Thus all the physical information is contained in two nondimensional parameters, $a$ and $\alpha=\frac{e^2}{\mu \omega^2 l^3}$. $a$ determines the distance between the harmonic oscillator and the Coulomb center, and depends only on the modulus of the magnetic momentum. $\alpha$ represents the ratio between the strengths of the Coulomb force and the harmonic oscillator Hooke force. From the point of view of the original problem, $\alpha$ depends only on the magnetic field strength. Solutions of the reduced system (\ref{redham}) will provide solutions for the original system by determining the dynamics of the center of mass using (\ref{CL1}) and (\ref{CL2}) and inverting the changes of variables, see Section 5.

\subsection{Separability}

We assume the $a\neq 0$ case, i.e. non-null magnetic momenta. The $a=0$ case requires a different approach and has been analyzed in \cite{Pinheiro,Turbiner}.

We introduce Euler elliptic coordinates relative to the foci: $f_1=(a,0)$ and $f_2=(-a,0)$, \cite{Perelomov}
\begin{eqnarray*}
&& q_1=a\, u v \ ,\  q_2^2= a^2\,  (u^2-1)(1-v^2) \quad ; \quad  -1 < v < 1 \ ,\  1 < u < \infty  \\ &&  \\
&& u=\frac{r_1+r_2}{2a} \  ,  \ v=\frac{r_2-r_1}{2a}   , \   r_1=\sqrt{(q_1-a)^2+q_2^2} \, \, , \, \, r_2=\sqrt{(q_1+a)^2+q_2^2}
\end{eqnarray*}

The reduced Hamiltonian (\ref{redham}) is written in these coordinates as:
\begin{equation}
{\cal H}=\frac{1}{2\mu a^2 (u^2-v^2)} \left[ (u^2-1) p_u^2+(1-v^2) p_v^2\right] + \frac{1}{u^2-v^2} \left[ f(u) + g(v) \right]\label{hEuler}
\end{equation}
where:
\begin{equation}
 f(u) = \frac{1}{2} a^2 (u^4-u^2) -\frac{\alpha}{a} u \  ,\quad
 g(v) = - \frac{1}{2} a^2 (v^4-v^2)+ \frac{\alpha}{a} v \nonumber \label{fyg}
\end{equation}
(\ref{hEuler}) is thus a Liouville Type I Hamiltonian \cite{Liouville,Perelomov}, and consequently admits two first integrals in involution given by \cite{Perelomov}:
\begin{equation}
{\cal H} = \frac{1}{u^2-v^2} \left[ {\cal H}_u + {\cal H}_v \right]  \  ,\quad
\Lambda = \frac{1}{u^2-v^2} \left[ u^2 {\cal H}_v +  v^2 {\cal H}_u \right] \label{hamlambda}
\end{equation}
where ${\cal H}_u$ and ${\cal H}_v$ are defined as:
\begin{eqnarray*}
{\cal H}_u &=& \frac{a^2}{2(u^2-1)} \left( \frac{d u}{d \zeta}\right)^2 + \frac{1}{2}a^2 (u^4-u^2) -\frac{\alpha}{a} u \\
{\cal H}_v &=& \frac{a^2}{2(1-v^2)} \left( \frac{d v}{d \zeta}\right)^2 - \frac{1}{2}a^2 (v^4-v^2) +\frac{\alpha}{a} v
\end{eqnarray*}
in terms of the velocities with respect to the local time $d \zeta = \frac{d t}{u^2-v^2}$. Fixing the values for the constants of motion: ${\cal H}=h$ and $\Lambda=\lambda$, we obtain from (\ref{hamlambda}) a separated first-order ODE's system:
\begin{eqnarray}
\left( \frac{du}{d\zeta}\right)^2 &=& \frac{2(u^2-1)}{a^2}\   \left( -\lambda + h \, u^2 - \frac{a^2}{2}  \, (u^4-u^2) + \frac{\alpha}{a} u\right) \label{eq1o1} \\
\left( \frac{dv}{d\zeta}\right)^2 &=& \frac{2(1-v^2)}{a^2}\    \left( \lambda-h \, v^2 +  \frac{a^2}{2} \, (v^4-v^2) - \frac{\alpha}{a} v \right) \label{eq1o2}
\end{eqnarray}
and thus the original equations are reduced to quadratures that depend on a physical parameter $\alpha$ and three constants of motion $a$, $h$ and $\lambda$.

Explicit solutions of (\ref{eq1o1}) and (\ref{eq1o2}) in terms of local time $\zeta$ would be obtained by inversion of hyperelliptic integrals of genus 2. This task is strongly complicated unlike other separable systems where one can integrate the equations in terms of elliptic functions. Nevertheless the study of the bifurcation diagram associated to (\ref{eq1o1},\ref{eq1o2}) is doable in the same way to the genus one case, as for instance the two Coulomb centers problem, see \cite{Waalkens,Seri,Gonzalez} and references therein.


\section{Bifurcation Diagram}

The analysis of the bifurcation diagram associated to equations (\ref{eq1o1}, \ref{eq1o2})  will allow us to classify the different types of orbits for the reduced system. This classification will be consequently inherited by the different types of motions of the original system. It is adequate to remark that each non trivial election of $x_0$ and $y_0$ in the original problem, leads to a reduced system characterized by $a=\sqrt{x_0^2+y_0^2}>0$. For simplicity we re-define the parameters appearing in (\ref{eq1o1}) and (\ref{eq1o2}) as:
\begin{equation}
h_a = \frac{h}{a^2} \quad , \quad \lambda_a=\frac{\lambda}{a^2} \quad , \quad \alpha_a = \frac{\alpha}{a^3} \nonumber\label{par}
\end{equation}
in such a way that these equations now read:
\begin{eqnarray}
\left( \frac{du}{d\zeta}\right)^2 &=& (1-u^2) (u^4 - (1 + 2 h_a) u^2 - 2 \alpha_a u + 2 \lambda_a) \label{neq1o1} \\
\left( \frac{dv}{d\zeta}\right)^2 &=& (1-v^2) (v^4 -(1 + 2 h_a) v^2 - 2 \alpha_a v + 2 \lambda_a)  \label{neq1o2}
\end{eqnarray}
and the dependence on $a$ is absorbed in the new parameters.

\begin{figure}
\begin{center}
  \includegraphics[height=6.5cm]{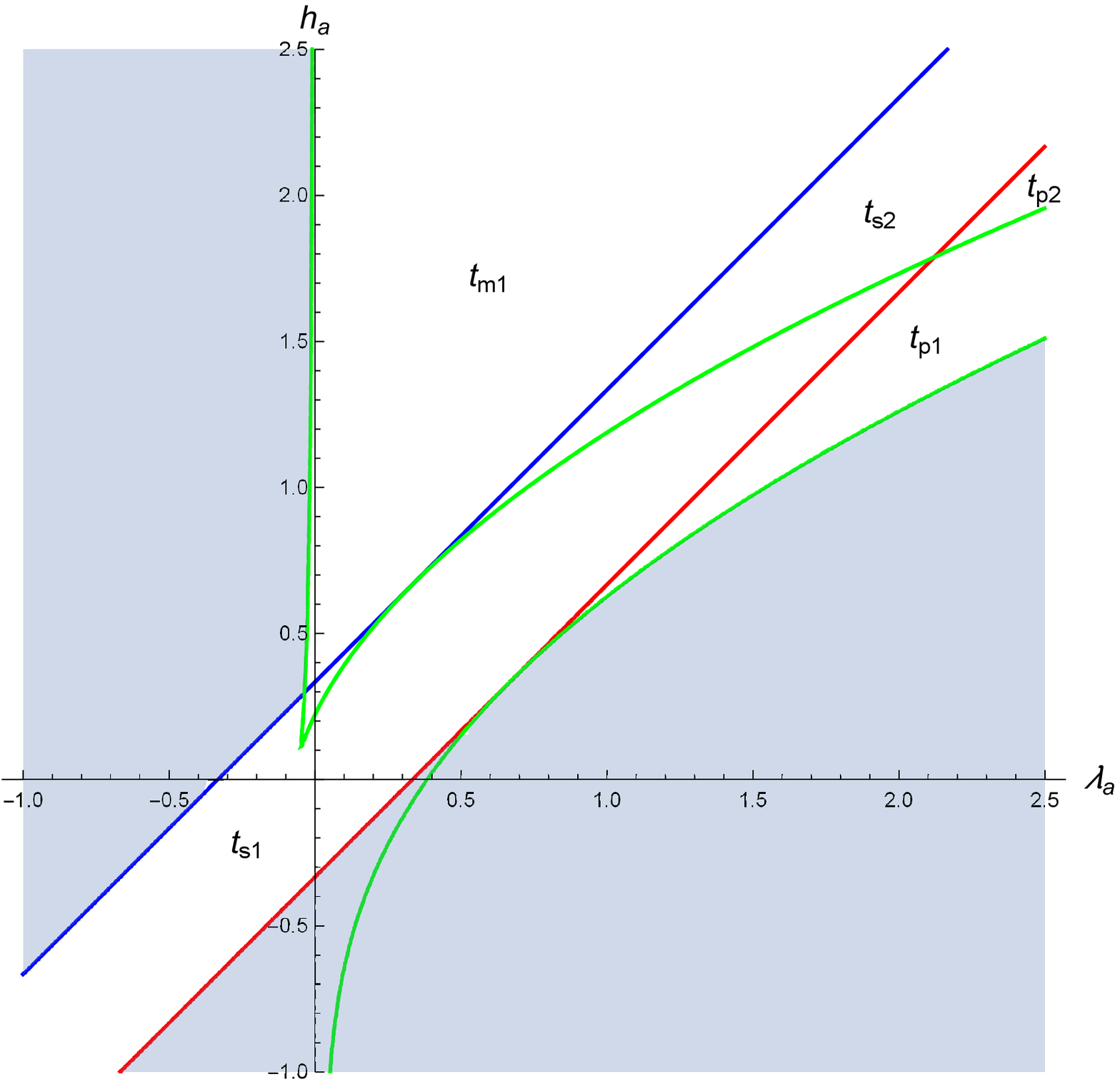} \qquad   \includegraphics[height=5cm]{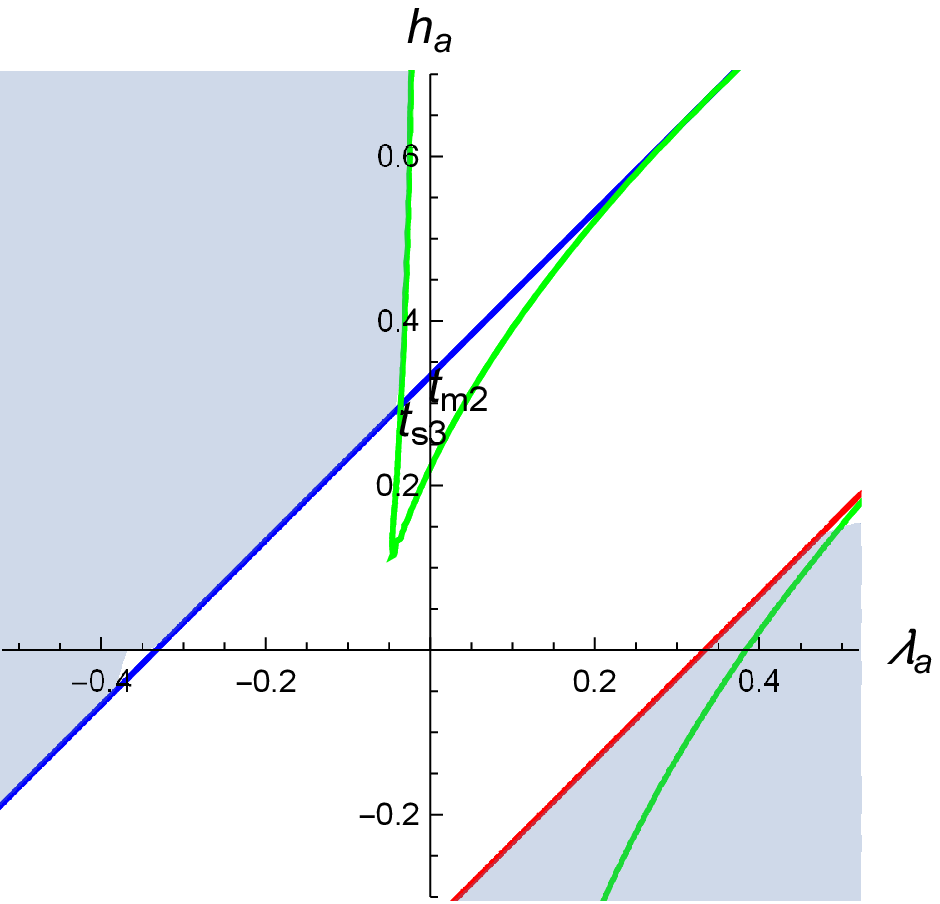}
\caption{Left) Bifurcation Diagram for: $\alpha_a= \frac{1}{3}$. Right) Detail for $t_{m2}$ and $t_{s3}$ zone.}
\label{bifurdia}       
\end{center}
\end{figure}

Both equations are determined by the same polynomial $P_6(z)= (1-z^2) P_4(z)$, with: $P_4(z)= z^4-(1+2h_a) z^2-2 \alpha_a z+2\lambda_a$, and $z=u$ or $z=v$ respectively in equation (\ref{neq1o1}) or (\ref{neq1o2}). Thus there are two fixed roots:  $z=\pm 1$, and four movable roots that depend on the parameter $\alpha_a$ and the constants of motion $h_a$ and $\lambda_a$.

For a given value of $\alpha_a$, the discriminant of $P_4(z)$:
\[
\Delta= 16 \left[ \lambda_a\,  (1+2 h_a) \, \left( (1+2 h_a)^2-72 \lambda_a) \alpha_a^2 + 2 (1+2 h_a)^2 -8 \lambda_a\right)^2  -27 \alpha_a^4 \right]
 \]
 determines the curve $\Delta=0$ on the $(\lambda_a,h_a)$ plane where at least two roots of $P_4(z)$ are identical. Besides this curve, we have to consider the straight lines: $h_a+ \alpha_a-\lambda_a=0$ and $h_a- \alpha_a-\lambda_a=0$, given by: $P_4(1)=0$ and $P_4(-1)=0$, where a root of $P_4(z)$ coincides with $+1$ or $-1$.

The bifurcation diagram corresponding to $\alpha_a=\frac{1}{3}$ is represented in Figure \ref{bifurdia}, Left), where the following color code is used:

\begin{itemize}

\item Red line: ${\cal L}_{\alpha_a}^1:=\{ h_a+ \alpha_a-\lambda_a=0\}$,

\item Blue line: ${\cal L}_{\alpha_a}^2:=\{ h_a - \alpha_a-\lambda_a=0\}$,

\item The green curves ${\cal L}_{\alpha_a}^3$ correspond with the implicit equation $\Delta=0$.

\item Shadowed areas represent regions where the motion is classically forbidden because $P_6(u)$ and/or $P_6(v)$ become negative.

\end{itemize}

There are six different allowed regions where the roots of $P_6(u)$ and $P_6(v)$ are simple, see Fig. \ref{bifurdia}. They can be classified in three types that we name as Satellitary, Planetary and Deformed oscillatory according to their qualitative behaviour.

\subsection*{Satellitary orbits}

This type of orbits is characterized by the presence of two caustic curves, ellipse and hyperbola, that bound the allowable motions for the reduced problem (shadowed regions in Fig. \ref{satreg}). The range of elliptic coordinates for this kind of orbits is: $1<u<u_c$ and $-1<v<v_c$, in such a way that:
\begin{eqnarray}
u = u_c &\Longrightarrow & \frac{q_1^2}{a^2 u_c^2} + \frac{q_2^2}{a^2 (u_c^2-1)}=1\label{ucaustic}\\
v = v_c &\Longrightarrow & \frac{q_1^2}{a^2 v_c^2} - \frac{q_2^2}{a^2 (1-v_c^2)}=1\label{vcaustic}
\end{eqnarray}
are the cartesian equations of the caustics in the $(q_1,q_2)$ plane, see Fig. \ref{satreg}. This situation appears in the bifurcation diagram in three different regions according with the relative disposition for the movable roots of $P_6(u)$ and $P_6(v)$ respectively:

\begin{figure}
\begin{center}
  \includegraphics[height=4cm]{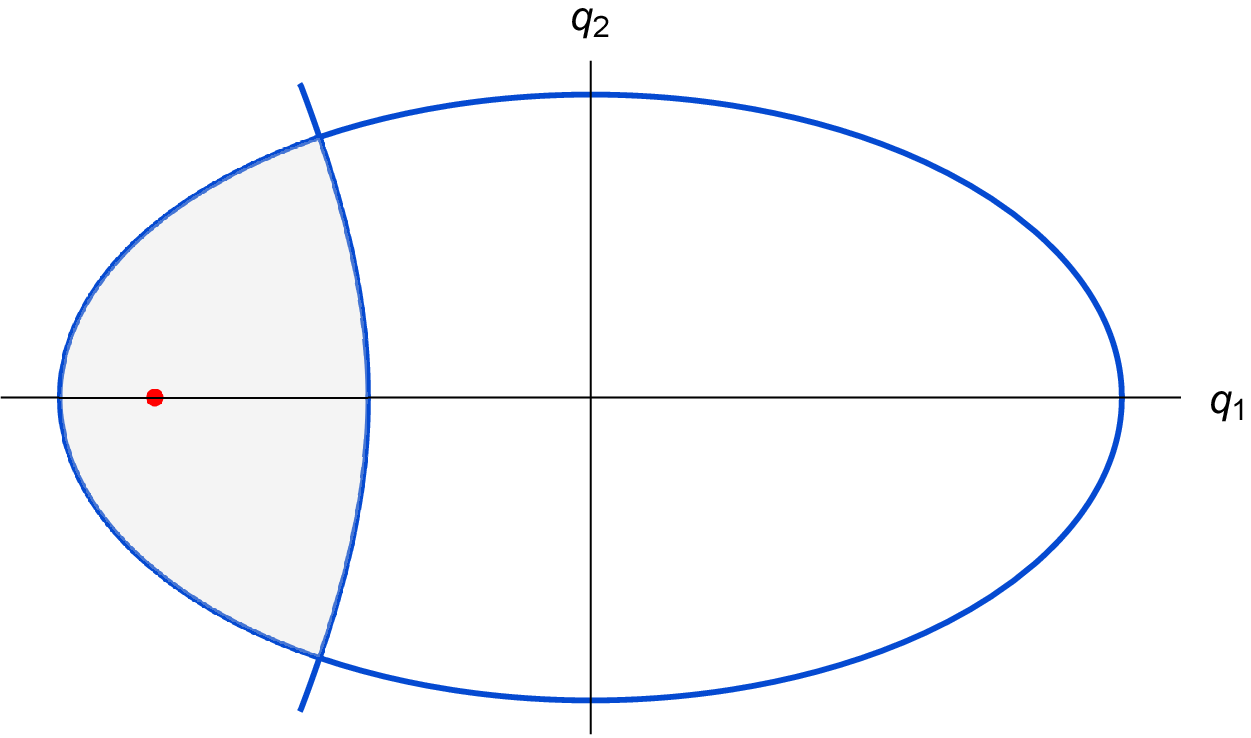} \qquad   \includegraphics[height=4.5cm]{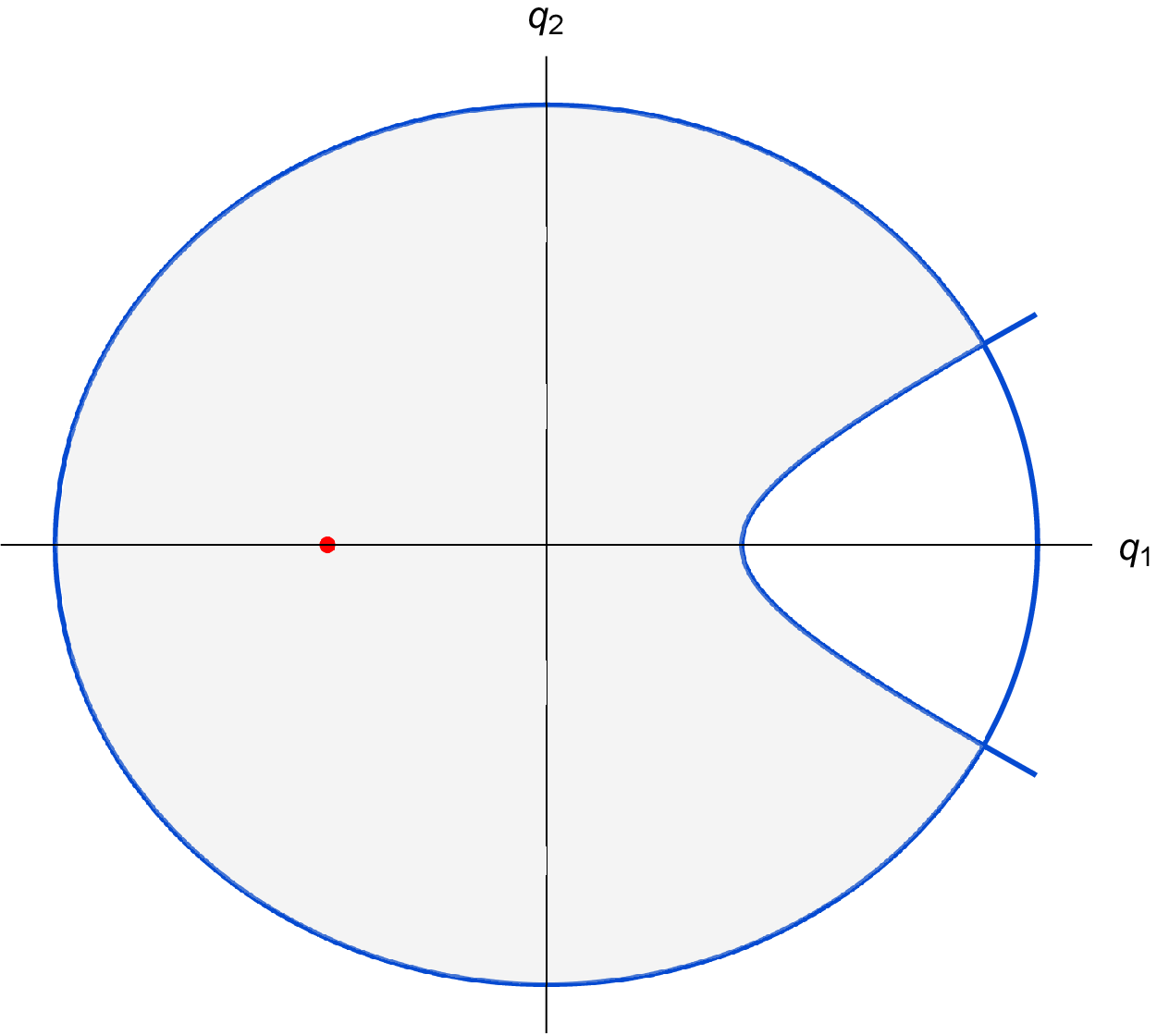}
\caption{Two examples of satellitary regions (shadowed) in the $(q_1,q_2)$ plane: $t_{s1}$ (left) and $t_{s2}$ (right). The harmonic oscillator is located at the origin and the red point represents the Coulomb center at $(-a,0)$.}
\label{satreg}       
\end{center}
\end{figure}

\begin{itemize}

\item $t_{s1}$ orbits:
\begin{equation}
\left\{  \begin{array}{l}  -1 < u_3 <  \mathbf{1 < u < u_4} \ , \ u_1, u_2 \in \mathbb{C} \\
\mathbf{ -1 < v < v_3} < 1 < v_4 \ , \ v_1, v_2 \in \mathbb{C}  \end{array} \right.
 \nonumber \label{ts1}
\end{equation}

\item $t_{s2}$ orbits:
\begin{equation}
\left\{  \begin{array}{l}  u_1 < u_2 < -1 < u_3 < \mathbf{1 < u < u_4}  \\
v_1 < v_2 < \mathbf{ -1 < v < v_3}  < 1 < v_4 \end{array} \right.
 \nonumber \label{ts2}
\end{equation}

\item $t_{s3}$ orbits:
\begin{equation}
\left\{  \begin{array}{l}  -1 < u_1 < u_2  < u_3 < \mathbf{ 1 < u < u_4}    \\
\mathbf{ -1 < v < v_1} < v_2  < v_3 < 1 < v_4  \end{array} \right.
  \nonumber\label{ts3}
\end{equation}

\end{itemize}

\subsection*{Planetary orbits}

\begin{figure}
\begin{center}
  \includegraphics[height=4cm]{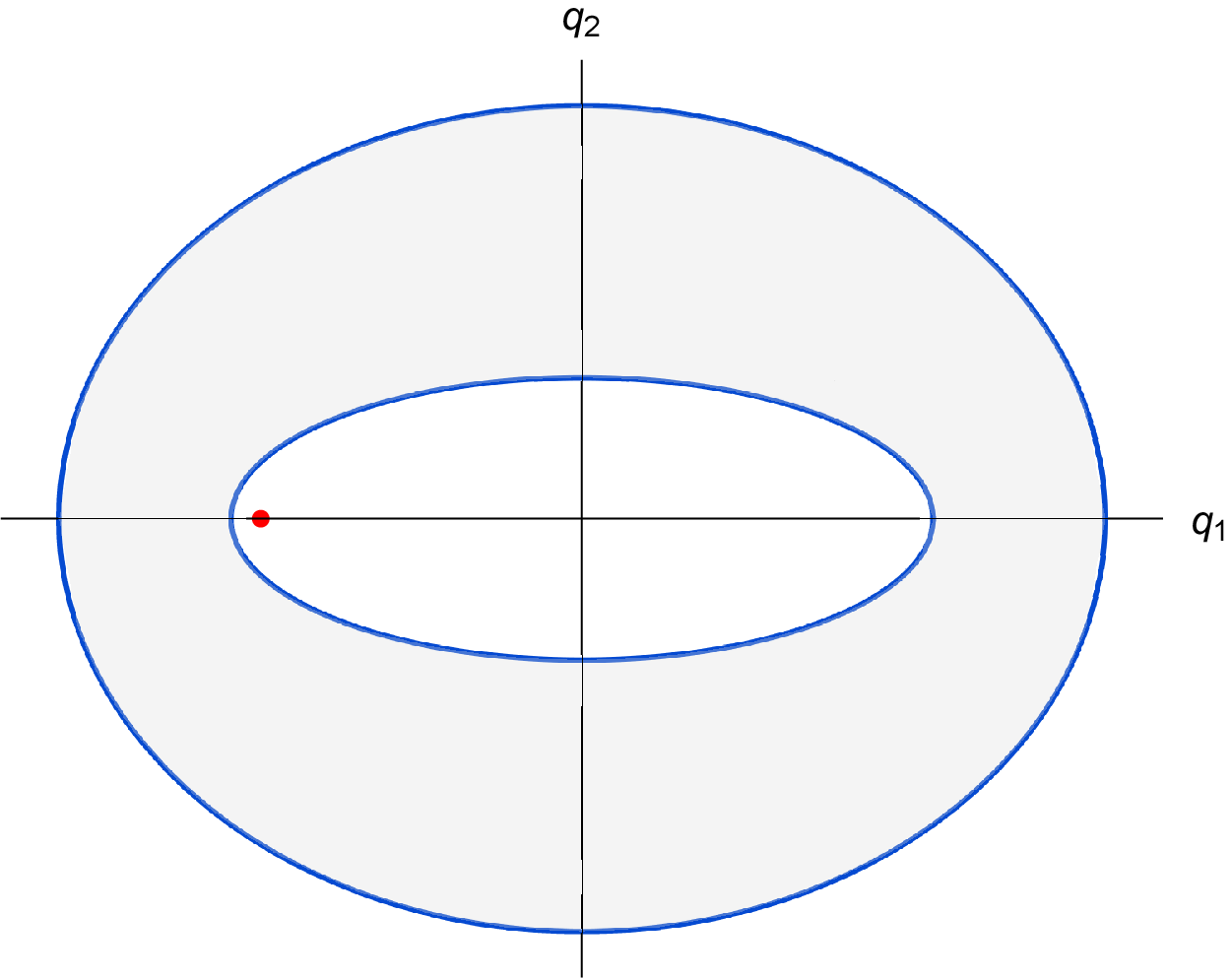} \qquad \includegraphics[height=4cm]{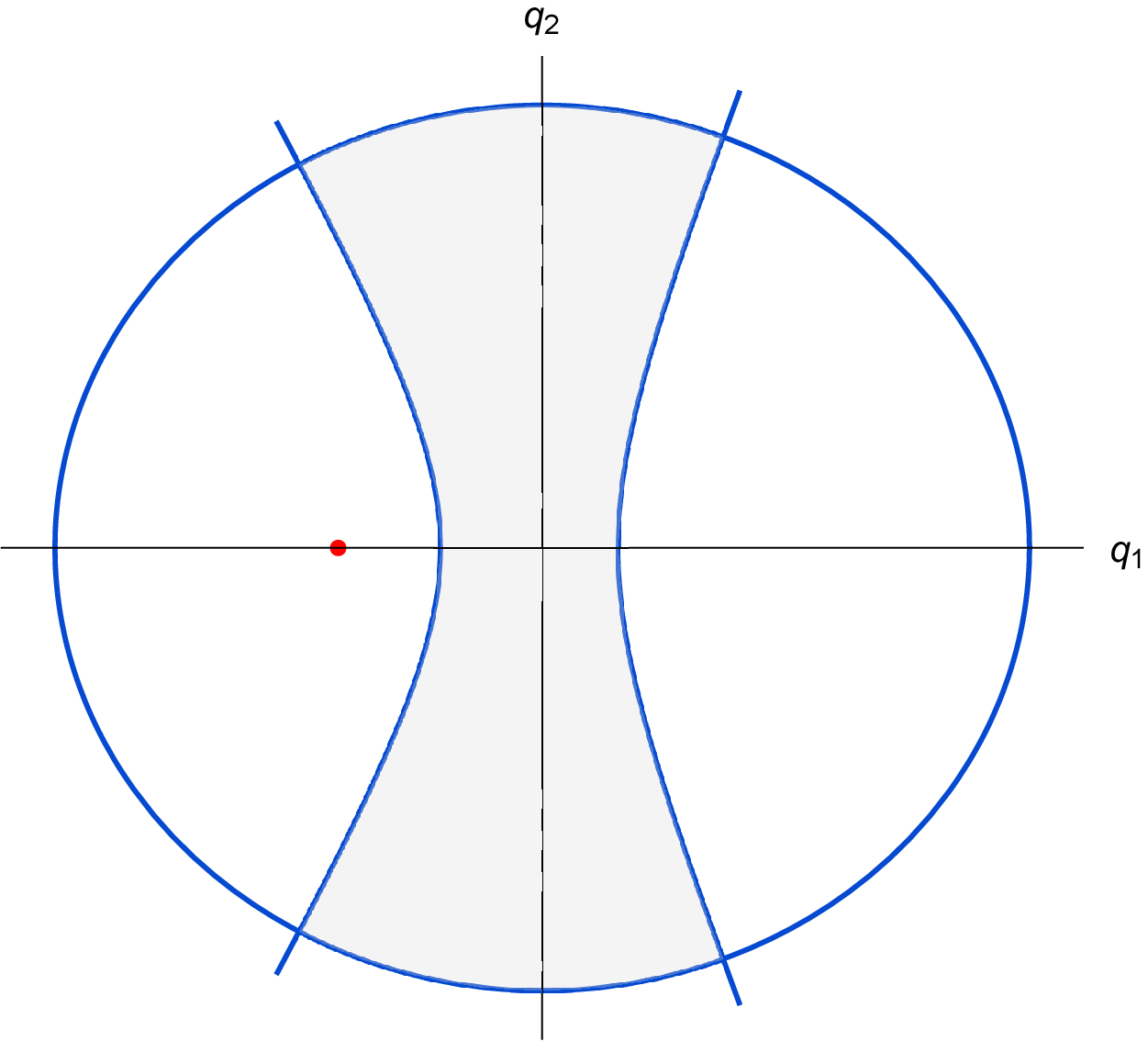}
  \caption{Two example of allowable regions in the $(q_1,q_2)$ plane: Left) Planetary orbits, Right) Deformed oscillatory orbits.}
\label{planreg}       
\end{center}
\end{figure}

In this case, the allowable motions are bounded by two ellipses: $u=u_{c1}$ and $u=u_{c2}$, i.e.: $1<u_{c1}< u <u_{c2}$, that in cartesian coordinates are written as (\ref{ucaustic}).  There are not restrictions in the $v$-coordinate, see Fig. \ref{planreg}, Left).

There are two different situations in the bifurcation diagram where this kind of orbits appear:

\begin{itemize}
\item $t_{p1}$ orbits:
\begin{equation}
\left\{  \begin{array}{l}  -1 < 1 < \mathbf{u_3 < u < u_4}  \ , \ u_1, u_2 \in \mathbb{C}    \\
-1 < v  < 1 < v_3 < v_4 \ , \ v_1, v_2 \in \mathbb{C}  \end{array} \right.
 \nonumber \label{tp1}
\end{equation}

\item $t_{p2}$ orbits:
\begin{equation}
\left\{  \begin{array}{l}  u_1 < u_2 < -1  < 1 < \mathbf{u_3 < u < u_4}   \\
v_1 < v_2 < -1 < v < 1 < v_3 < v_4 \end{array} \right.
\nonumber  \label{tp2}
\end{equation}

\end{itemize}

\subsection*{Deformed oscillatory orbits}

The caustics are now two hyperbolas: $v=v_{c1}$ and $v=v_{c2}$, with: $-1<v_{c1}<v<v_{c2}<1$, and the ellipse: $u=u_{c}$, i.e. $1<u<u_c$, see Fig. \ref{planreg} Right). The corresponding equations in the $(q_1,q_2)$ plane read as (\ref{vcaustic}) and (\ref{ucaustic}) respectively. Two regions of the bifurcation diagram give rise to this type of orbits:

\begin{itemize}

\item $t_{m1}$ orbits:
\begin{equation}
\left\{  \begin{array}{l}  u_1 < -1 < u_2  < u_3 < \mathbf{1 < u < u_4} \\
 v_1 < -1 < \mathbf{v_2 < v < v_3} < 1 < v_4 \end{array} \right.
 \nonumber \label{tm1}
\end{equation}

\item $t_{m2}$ orbits:
\begin{equation}
\left\{  \begin{array}{l}  -1 < u_1 < u_2  < u_3 < \mathbf{1 < u < u_4} \\
 -1 < v_1 < \mathbf{v_2 < v < v_3} < 1 < v_4\end{array} \right.
 \nonumber \label{tm2}
\end{equation}

\end{itemize}

\begin{figure}
\begin{center}
  \includegraphics[height=4cm]{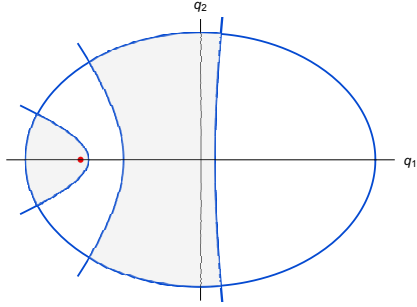}
  \caption{Allowable regions for motion of the relative particle in the $(q_1,q_2)$ plane for $t_{s3}$ and $t_{m2}$ orbits corresponding to a given pair of values of $h_a$ and $\lambda_a$.}
\label{mixta}       
\end{center}
\end{figure}

It has to be remarked that $t_{s3}$ satellitary and $t_{m2}$ deformed oscillatory orbits correspond with the same region in the bifurcation diagram, i.e. the area limited by the discriminant curve and the straight line: $h_a-\alpha_a-\lambda_a=0$, see Fig. \ref{bifurdia} Right). Thus, for given values of $h_a$ and $\lambda_a$ in this region, the relative particle will travel on one type of orbit or another depending on its initial position inside an allowable satellitary or oscillatory region respectively, see Fig. \ref{mixta}.

It is interesting to note that all the allowable motions for this reduced problem are bounded regardless the sign of the mechanical energy $h_a$ and the value of the second constant of motion $\lambda_a$, no unbounded motions are possible in the reduced problem.


\section{Electron-positron motions}

The classification of orbits performed in Section 4 can be translated one-to-one to the original problem of an electron-positron pair in a constant orthogonal magnetic field if the constants of motion $K_X$ and $K_Y$, thus $x_0$ and $y_0$, have been fixed for non simultaneously null values. Moreover, the explicit knowledge of $x(t)$ and $y(t)$, i.e. the solutions of the reduced system, would lead to a direct calculation for the center of mass dynamics by integration of equations (\ref{CL1}, \ref{CL2}), that in non-dimensional variables read:
\begin{equation}
\dot{X}=\frac{1}{2} (y+y_0)\ ; \quad  \dot{Y}=\frac{1}{2} (-x+x_0)\label{nd-CL}
\end{equation}

However, the obtention of explicit expressions for the original physical problem trajectories  would require to accomplish several cumbersome steps: 1. Inversion of the hyperelliptic integrals arising from (\ref{eq1o1}) and (\ref{eq1o2}) to obtain the functions $u(\zeta)$ and $v(\zeta)$ in terms of the local time $\zeta$; 2. Integration and inversion of equation: $dt= (u^2(\zeta)-v^2(\zeta)) d\zeta$ in order to describe the dependence on the physical time $t$; 3. Writing the trajectory solutions $(x(t), y(t))$ of the reduced system, i.e. the relative motion for the original problem; 4. Integration of equations (\ref{nd-CL}) to determine the dynamics of the center of mass; and 5. Finally, it would be straightforward to write the explicit solutions $(x_1(t), y_1(t))$ and $(x_2(t), y_2(t))$ for the trajectories of the positron and electron respectively. It is therefore compulsory to consider a numerical approach to obtain in an efficient way the dynamics of the two particles.

Starting from fixed values of $K_X$ and $K_Y$, it is possible to numerically analyze equations (\ref{eq1o1}) and (\ref{eq1o2}) by considering initial conditions compatible with the allowable regions of the bifurcation diagram for the reduced problem. Nevertheless it is more convenient to approach directly the Newton equations (\ref{Newtoneqs}) of the original problem, that in dimensionless variables read:
\begin{equation}
\begin{array}{ll}
\displaystyle  \ddot{x}_1 = -  \dot{y}_1 - \frac{\alpha}{2r^3} (x_1-x_2), \quad &
\displaystyle  \ddot{y}_1 = \ \  \dot{x}_1 - \frac{\alpha}{2r^3} (y_1-y_2) \\ &\\
\displaystyle  \ddot{x}_2 = \ \  \dot{y}_2 + \frac{\alpha}{2r^3} (x_1-x_2),\quad &
\displaystyle \ddot{y}_2 = -  \dot{x}_2 + \frac{\alpha}{2r^3} (y_1-y_2) \end{array}
\label{NLnd}
\end{equation}
and obviously it is necessary to know the initial conditions for positions, $(x_i(0),y_i(0))$, and velocities, $(\dot{x}_i(0),\dot{y}_i(0))$, $i=1,2$. Compatibility with the obtained classification of orbits is guaranteed with the following procedure: Fixed values of $(x_0,y_0)$ specify a bifurcation diagram for the reduced system. We can then choose a pair of values of the constants of motion, ${\cal H}=h$ and $\Lambda=\lambda$, characterizing a concrete type of orbit and determining the corresponding allowable region in the $(q_1,q_2)$ plane for the motion of the relative particle. We must also to choose an initial position for the relative particle inside this region: $(q_1(0),q_2(0))$, and an initial position for the center of mass, i.e. $(X(0),Y(0))$.

Using the expressions (\ref{hamlambda}) of ${\cal H}$ and $\Lambda$ in Cartesian coordinates:
\begin{eqnarray}
  {\cal H} &=& \frac{1}{2} (\dot{q}_1^2+\dot{q}_2^2) + \frac{1}{2}  (q_1^2+q_2^2) -\frac{\alpha}{\sqrt{(q_1+a)^2+q_2^2}} \label{hnd1} \\
   \Lambda &=& \frac{1}{2} (L^2 + \dot{q}_1^2) + \frac{1}{2}  q_1^2 + \frac{\alpha q_1}{a \sqrt{(q_1+a)^2+q_2^2}} \label{sind2}
\end{eqnarray}
with $L=q_2 \dot{q}_1-q_1 \dot{q}_2$, one can obtain the associated values of $(\dot{q}_1(0),\dot{q}_2(0))$ in terms of $h$ and $\lambda$. Having into account that the dependence on velocities in (\ref{hnd1}) and (\ref{sind2}) is quadratic, the correspondence $(h,\lambda)$ to  $(\dot{q}_1,\dot{q}_2)$ is one-to-four. Finally, (\ref{nd-CL}) and the inversion of the changes of variables lead us to the following expressions of the initial conditions for equations (\ref{NLnd}):
\begin{eqnarray*}
x_1(0) &=& X(0) - \frac{1}{2a} \left( x_0 (a +q_1(0)) +y_0 q_2(0)\right)  \\
y_1(0) &=& Y(0) + \frac{1}{2a} \left( y_0 (a+q_1(0)) - x_0 q_2(0)\right) \\
x_2(0) &=& X(0) + \frac{1}{2a} \left( x_0 (a +q_1(0)) +y_0 q_2(0)\right)  \\
y_2(0) &=& Y(0) - \frac{1}{2a} \left( y_0 (a+q_1(0)) - x_0 q_2(0)\right)
\end{eqnarray*}
and
\begin{eqnarray*}
\dot{x}_1(0) &=& \  \  \frac{1}{2a} \left[ x_0 (q_2(0) - \dot{q}_1(0)) - y_0 ( q_1(0)+ \dot{q}_2(0) )\right] \\
\dot{y}_1(0) &=& - \frac{1}{2a} \left[ x_0 (q_1(0) + \dot{q}_2(0)) + y_0 (q_2(0)-\dot{q}_1(0) )\right] \\
\dot{x}_2(0) &=& \ \  \frac{1}{2a} \left[ x_0 (q_2(0) + \dot{q}_1(0)) - y_0 (q_1(0)-\dot{q}_2(0))\right] \\
\dot{y}_2(0) &=& - \frac{1}{2a} \left[ x_0 (q_1(0) - \dot{q}_2(0)) + y_0 (q_2(0)+\dot{q}_1(0) )\right] \\
\end{eqnarray*}
depending only in $x_0 = \frac{K_2}{m\omega l}$, $y_0=\frac{K_1}{m\omega l}$, $q_1(0)$, $q_2(0)$, $\dot{q}_1(0)$, $\dot{q}_2(0)$ and $X(0)$, $Y(0)$. We will assume that $X(0)=0$ and $Y(0)=0$ without loss of generality.

\begin{figure}
\begin{center}
\includegraphics[height=3.9cm]{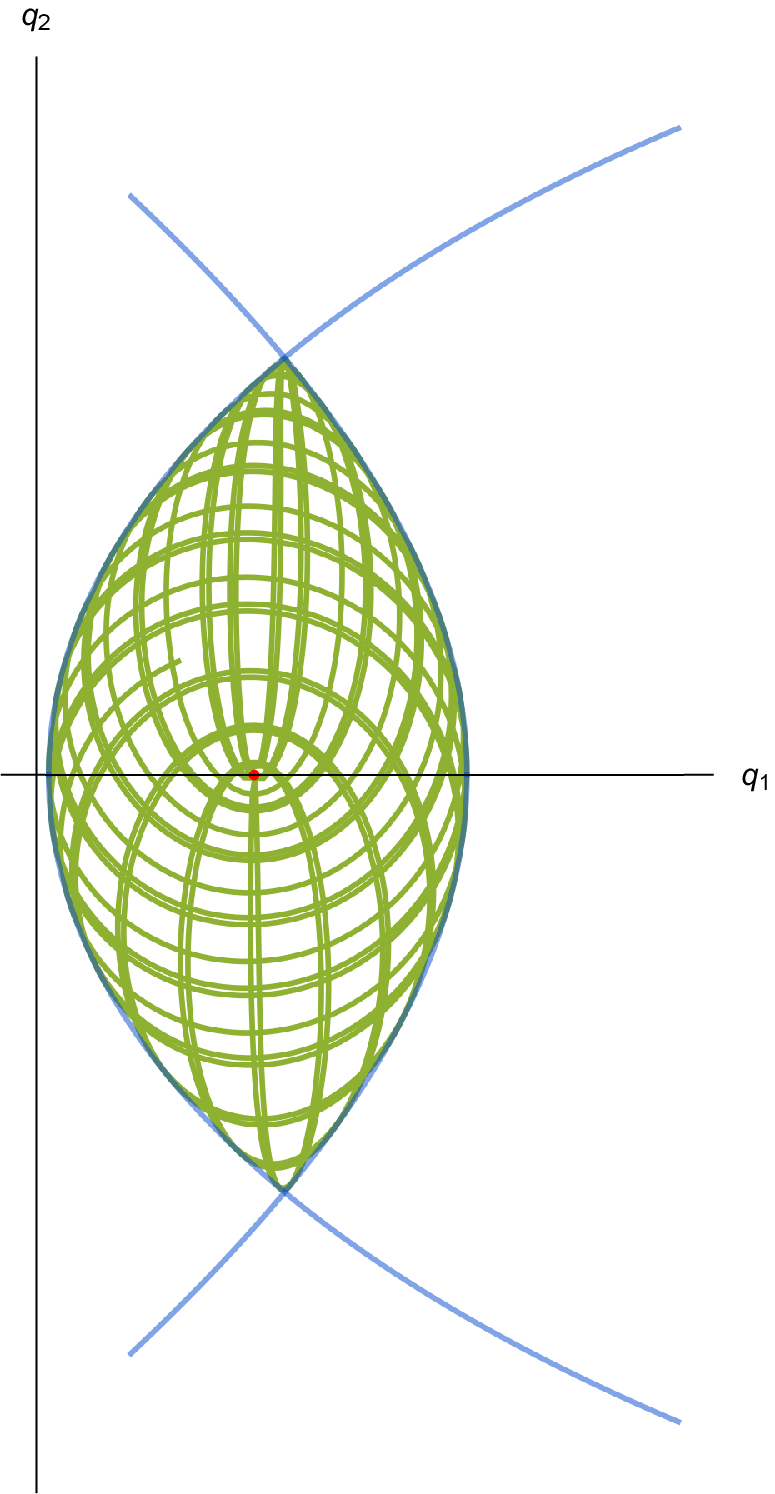}  \  \includegraphics[height=2.3cm]{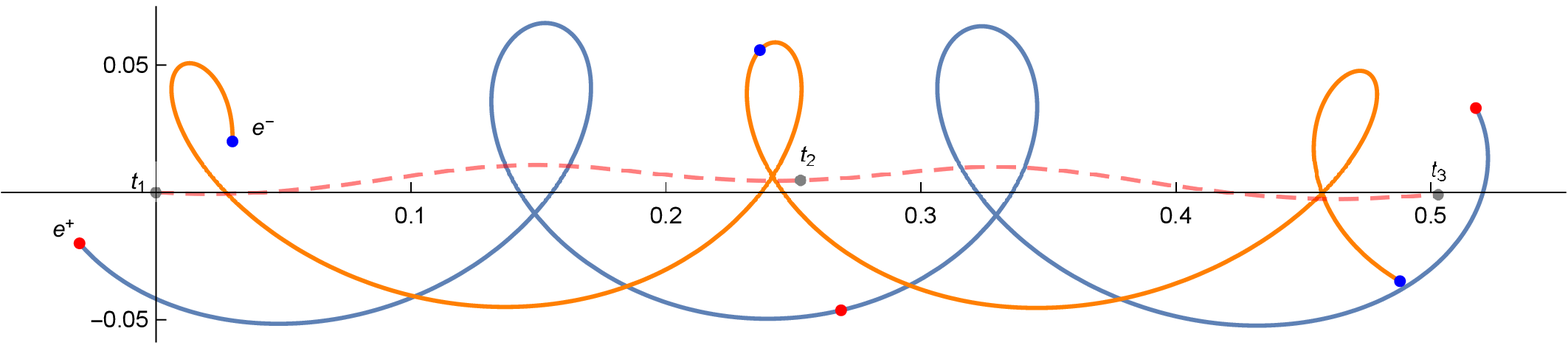}
\caption{$t_{s1}$ type orbit with: $\alpha_a=\frac{1}{3}$, $h_a= -1$, $\lambda_a=-1$, $(x_0,y_0)=(0,1)$, $q_1(0)=-1.04$, $q_2(0)=0.06$, $X(0)=Y(0)=0$. Left) Relative particle orbit with $t\in[0,10]$. Right) Electron(orange)-positron(blue) and center of mass (red-dashed) orbits for $t\in[0,1]$. Highlighted points correspond with the positions of the particles at the initial time: $t_1=0$, the intermediate time $t_2$ and the final instant $t_3$.}
\label{SatExample1}
\end{center}
\end{figure}

\subsection*{Description of the orbits}

The principal features for the three different types of orbits will be described in this section by means of selected examples.

A satellitary $t_{s1}$ orbit with $\alpha_a=\frac{1}{3}$, $h_a= -1$ and $\lambda_a=-1$ is represented in Fig. \ref{SatExample1}. The relative particle motion is bounded by the two caustic curves (\ref{ucaustic},\ref{vcaustic}) $u_c=1.108$ and $v_c=-0.887$ that determine an small region around the Coulomb center for this case with $h_a<0$, see Fig. \ref{SatExample1} Left). Assuming that the orbit is dense, i.e. the relative motion is not periodic, we conclude that the relative particle will end falling into the center, and thus the electron and positron will collide after a finite time. Figure \ref{SatExample1} has been drawn by choosing the initial conditions: $\dot{q}_1(0)=-2.28$, $\dot{q}_2(0)=-0.97$ between the four possibilities given by $h_a=\lambda_a=-1$ and $(q_1(0),q_2(0))=(-1.04,0.06)$ in equations (\ref{hnd1},\ref{sind2}). Fig. \ref{AppEj1} in the Appendix shows a second example of $t_{s1}$ orbit, here the value $h_a>0$ generates an allowable region considerably bigger for the relative particle motion.

Figures \ref{AppEj3} and \ref{AppEj2} in the Appendix show two examples of $t_{s3}$ and $t_{s2}$ orbits, with similar features to those in Fig. \ref{SatExample1} and Fig. \ref{AppEj1} respectively.

\begin{figure}
\begin{center}
\includegraphics[height=4.1cm]{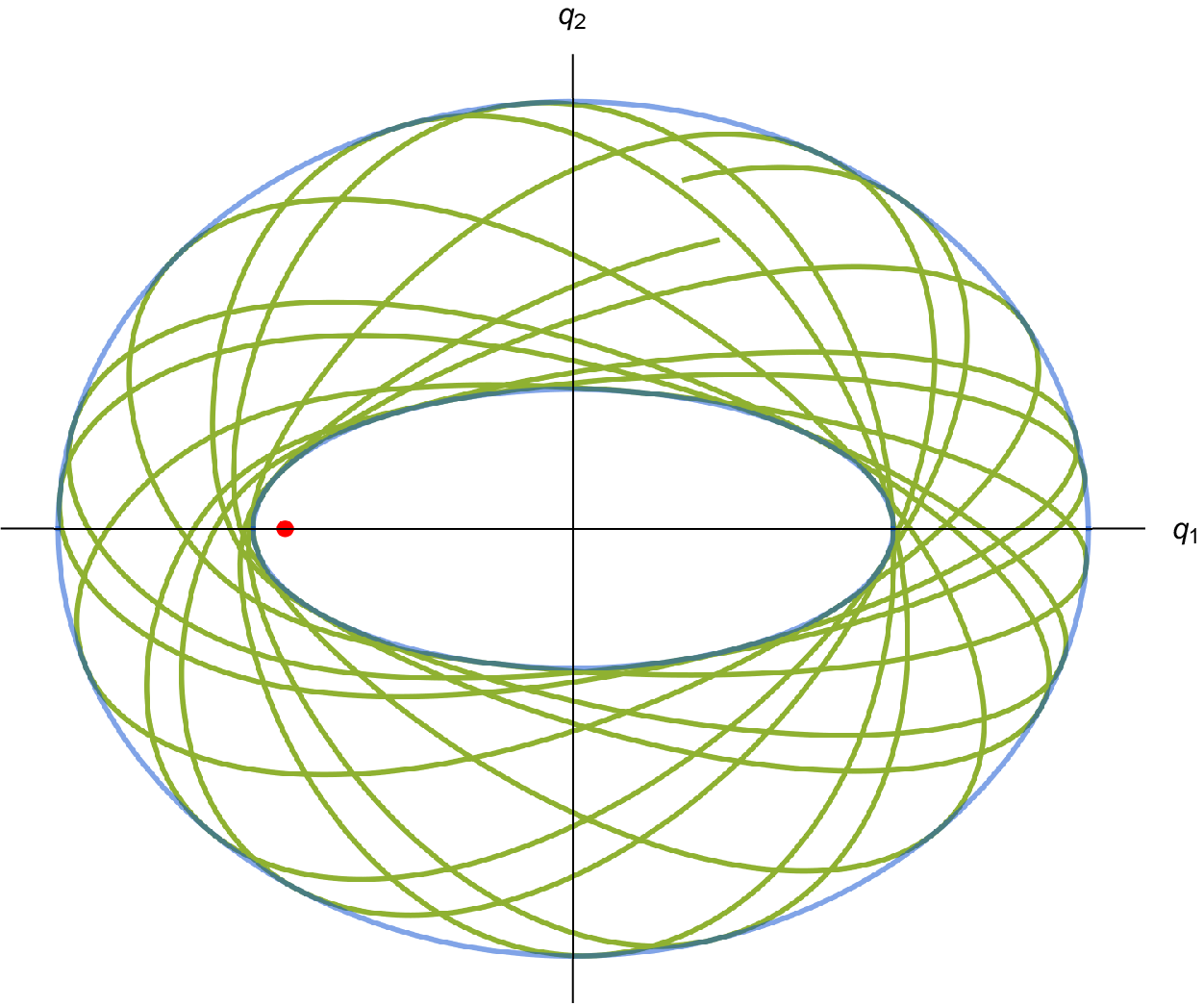}  \  \includegraphics[height=4.2cm]{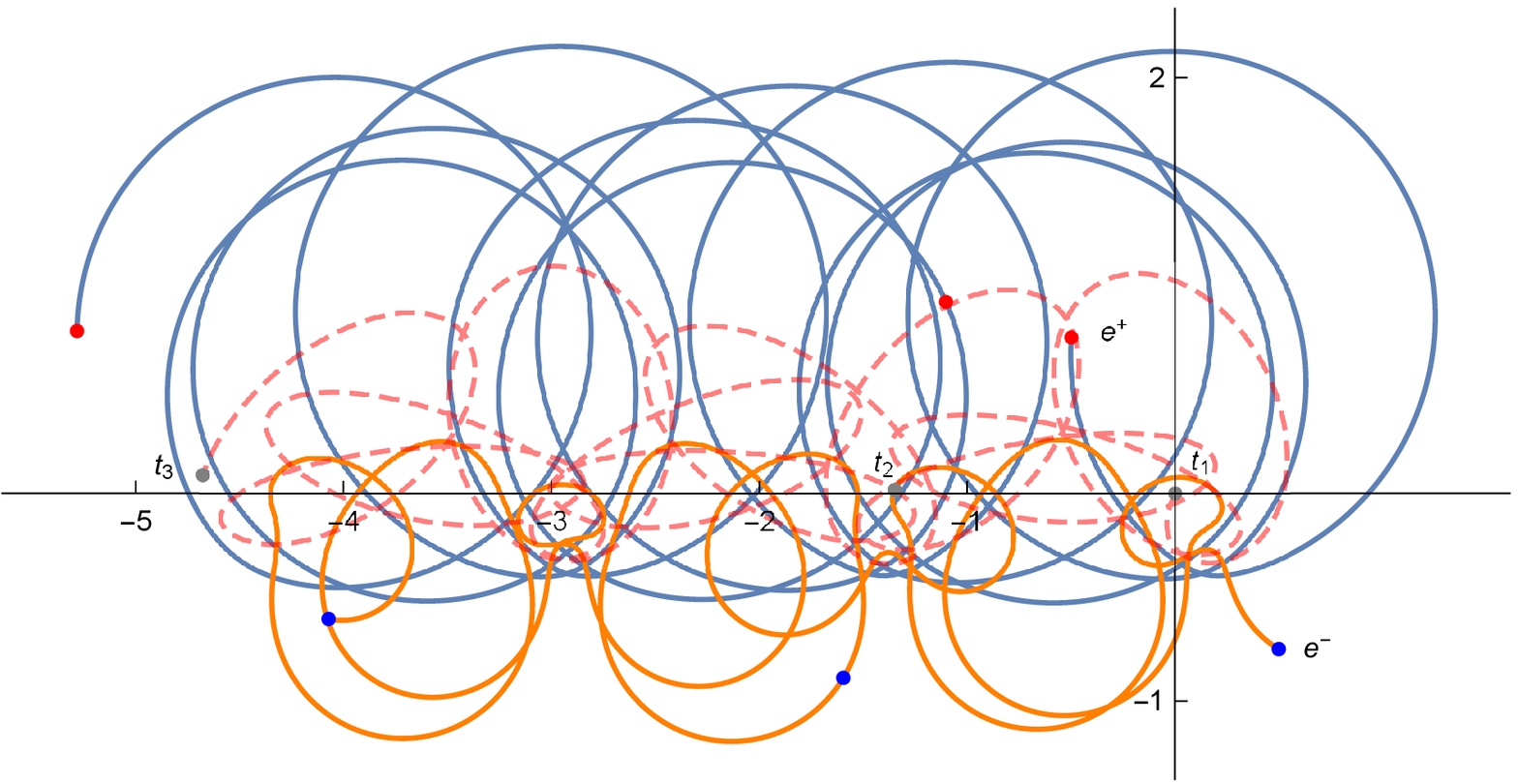}
\caption{Example of a $t_{p1}$ planetary orbit with: $\alpha_a=\frac{1}{3}$, $h_a=1.6$, $\lambda_a=2.2$, $(x_0,y_0)=(0,1)$, $q_1(0)=0.5$, $q_2(0)=1$, and $t\in[0,60]$. Left) Relative particle orbit. Right) Electron(orange)-positron(blue) orbits and center of mass trajectory (red-dashed). }
\label{PlanetaryEj1}
\end{center}
\end{figure}

Figure \ref{PlanetaryEj1} shows a $t_{p1}$ planetary orbit with $h_a=1.6$, $\lambda_a=2.2$ and the same choice for $(x_0,y_0)=(0,1)$. The allowable region for the relative particle does not include the Coulomb center, see Fig. \ref{PlanetaryEj1} Left), and consequently electron and positron will never collide.  The caustics for the relative motion are the ellipses $u_{c1}=1.111$ and $u_{c2}=1.788$. Fig. \ref{AppEj4} in Appendix is another planetary orbit, in this case of $t_{p2}$ type.

\begin{figure}
\begin{center}
\includegraphics[height=4.5cm]{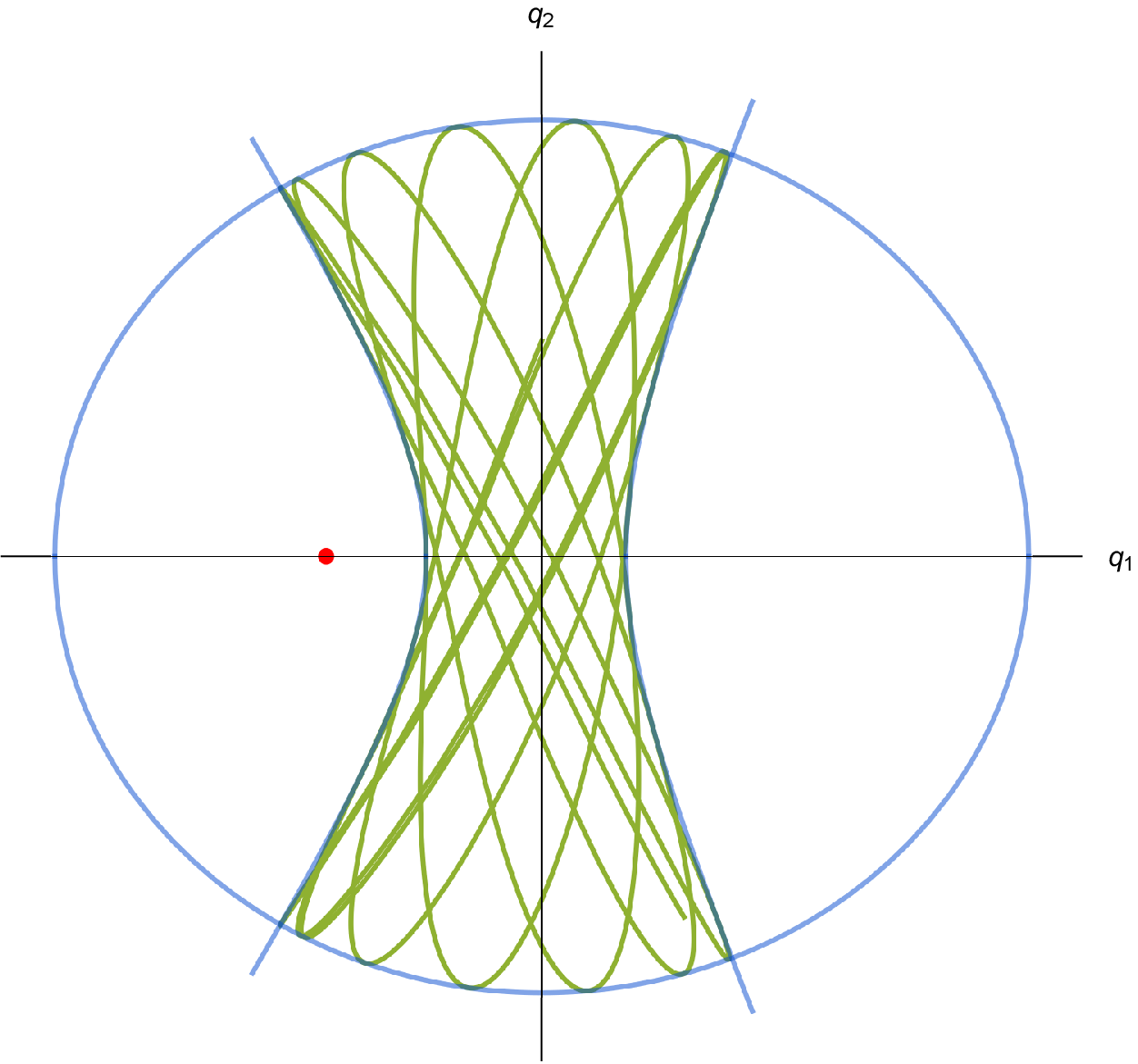} \   \includegraphics[height=4.8cm]{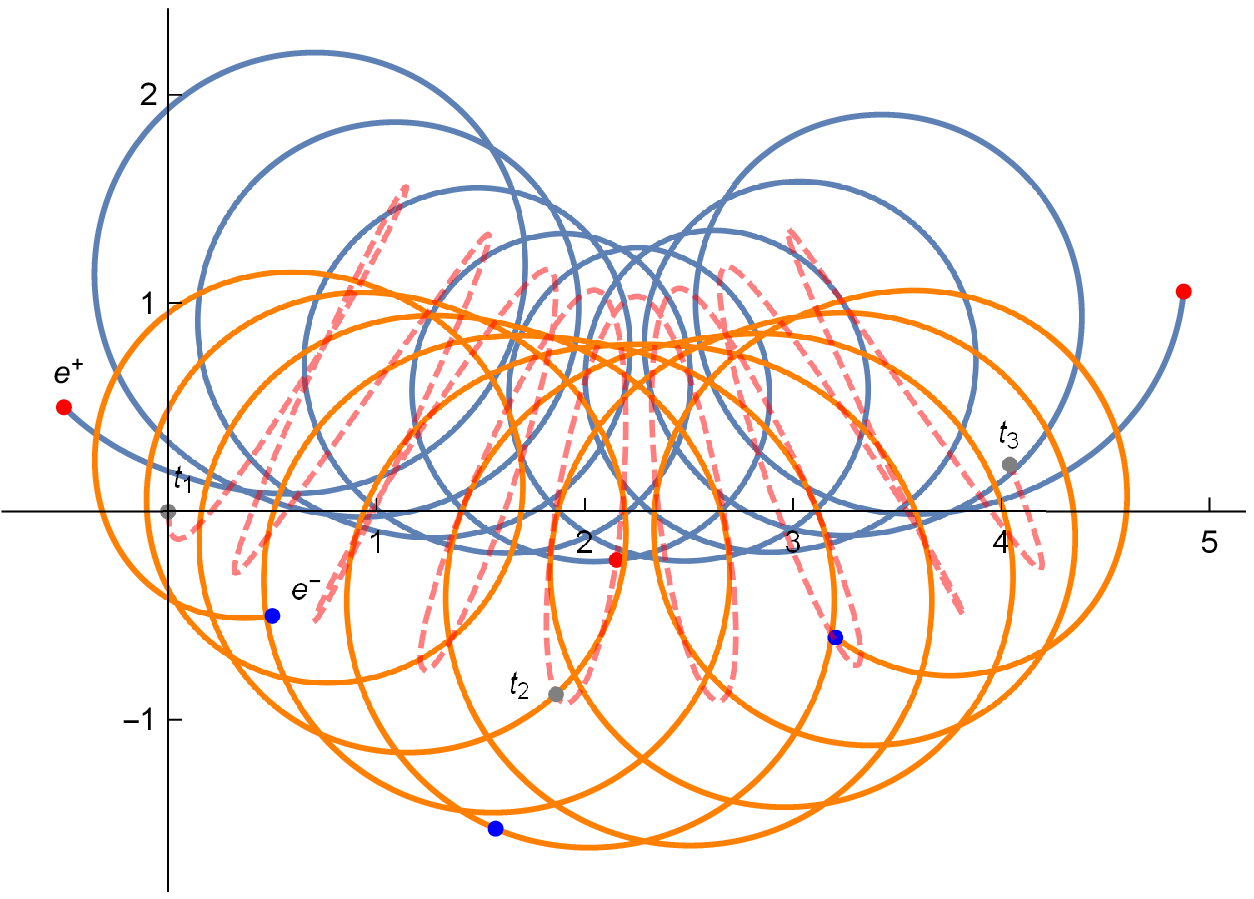}
\caption{Example of $t_{m1}$ oscillatory orbit with: $\alpha_a=\frac{1}{3}$, $h_a=2$, $\lambda_a=0.5$, $(x_0,y_0)=(0,1)$, $q_1(0)=0$, $q_2(0)=1$ and $t\in[0,50]$. Left) Relative particle. Right) Electron(orange)-positron(blue) orbits and center of mass trajectory (red-dashed). }
\label{OscillatoryEj1}
\end{center}
\end{figure}

Finally Fig. \ref{OscillatoryEj1} corresponds to a $t_{m1}$ deformed oscillatory orbit with: $\alpha_a=\frac{1}{3}$, $h_a=2$, $\lambda_a=0.5$, $(x_0,y_0)=(0,1)$. The caustics are in this case the ellipse $u_c=2.258$ and the hyperbolas: $v_{c1}=-0.537$, $v_{c2}=0.391$. We find again a non-collision situation because the allowable region does not contain the Coulomb center. Fig. \ref{AppEj5} is a $t_{m2}$ deformed oscillatory orbit.


The choice $(x_0,y_0)=(0,1)$ determine a privileged direction for the electron - positron motion. The magnetic momentum is proportional to $(y_0,x_0)$, i.e. $(K_1,K_2) \propto (1,0)$ and thus the motion of the pair of particles can be interpreted as a global displacement in the horizontal axis plus a local motion around this direction, see Figures \ref{SatExample1} Right), \ref{PlanetaryEj1} Right) or \ref{OscillatoryEj1} Right). We can see in Fig. \ref{AppEj7} the same phenomenon for other choices of $(x_0,y_0)$ and therefore different directions for the global translational motion.


Figures \ref{AppEj8} illustrate the previously commented fact that a concrete choice of $h_a, \lambda_a, q_1(0)$ and $q_2(0)$ lead in equations (\ref{hnd1},\ref{sind2}) to four possibilities for $(\dot{q}_1(0),\dot{q}_2(0))$.  The graphics in \ref{AppEj8} correspond to all these possibilities and reflect the symmetries with respect to the interchange between electron and positron plus spatial inversion.

Finally, it is interesting to remark that the value $\alpha=\frac{1}{3}<1$ implies that the Coulomb center strength is weaker than the harmonic oscillator one, thus the magnetic field of the original problem is stronger than the electrical interaction between the two particles. An inverse situation, with $\alpha=2>1$, is shown in Fig. \ref{AppEj6}. We observe that the global translational displacement is considerably more relevant that the local motion for this situation where the Coulomb interaction is dominant compared with the magnetic strength.

\section{Further comments}

Future developments for this work would include the inversion of equations (\ref{eq1o1}) and (\ref{eq1o2}) in terms of hyperelliptic $\theta-$funtions. Explicit analytical solutions would permit the study of periodic solutions for the relative particle. Having accomplished this task, it would be possible to approach the WKB method for the quantum problem.

Another possible field of study is the analysis of the reduced problem in the $S^2$-sphere. Using projective techniques, see \cite{ProVer} and references therein, this system will be separable in sphero-conical coordinates. The interesting related problem of two centers in $S^2$ in the presence of the Dirac magnetic monopole has been recently worked out in \cite{Veselov}.

\section*{Acknowledgements}

The authors warmly acknowledge M Gadella for illuminating conversations on several issues concerning this work. We also thank the Spanish Ministerio de Econom\'{\i}a y Competitividad (MINECO) and the Junta de Castilla y Le\'on for partial financial support under grants MTM2014-57129-C2-1-P, and VA057U16, BU229P18 and SA067G19.



\appendix

\section{Gallery of orbits}



\begin{figure}[h]
\begin{center}
\includegraphics[height=4.2cm]{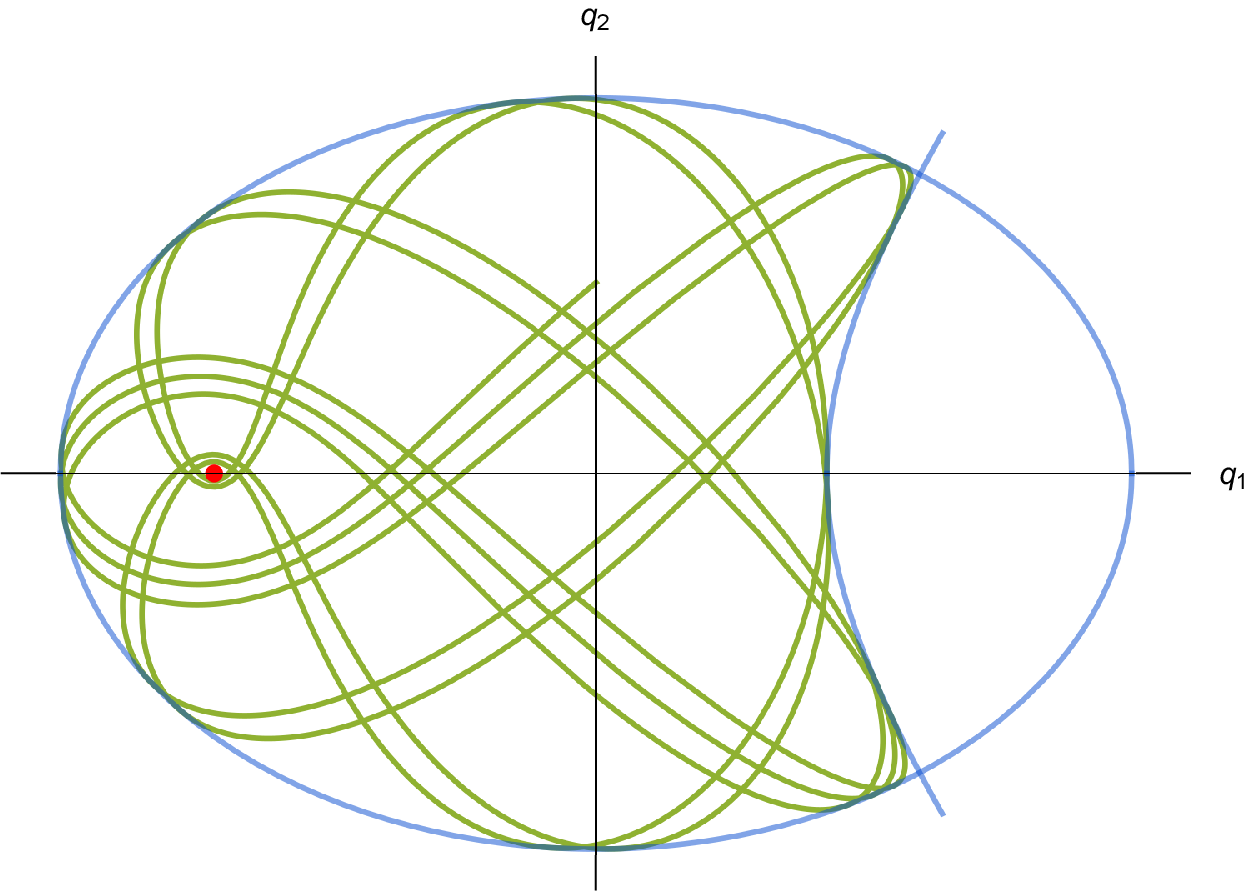} \   \includegraphics[height=5.2cm]{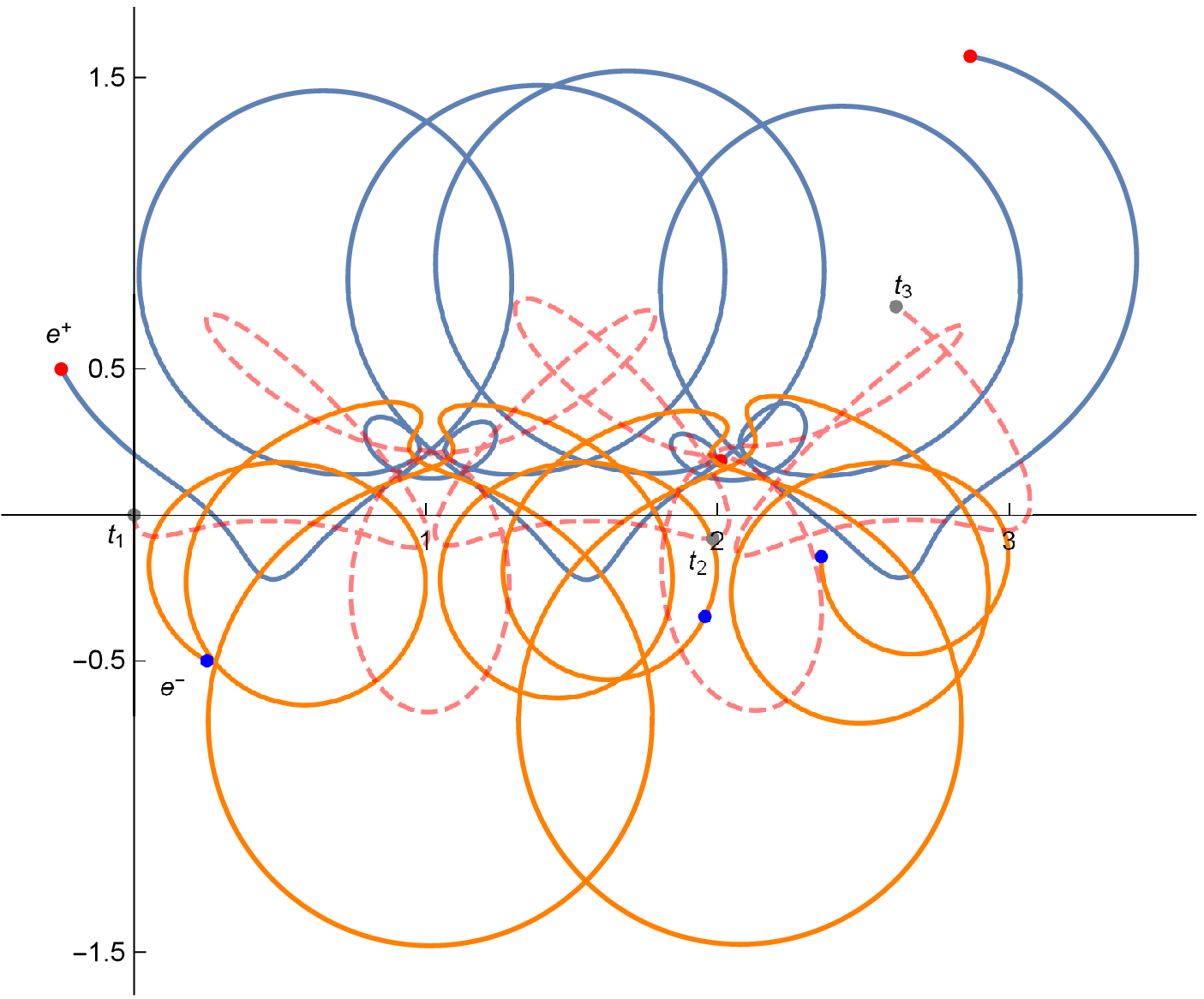}
\caption{$t_{s1}$ orbit with: $\alpha_a=\frac{1}{3}$, $h_a=0.5$, $\lambda_a=0.5$, $(x_0,y_0)=(0,1)$, $q_1(0)=0$, $q_2(0)=0.5$, and $t\in[0,40]$. Left) Relative particle orbit. Right) Electron(orange)-positron(blue) orbits and center of mass trajectory (red-dashed).}
\label{AppEj1}
\end{center}      
\end{figure}


\begin{figure}[h]
\begin{center}
\includegraphics[height=3.3cm]{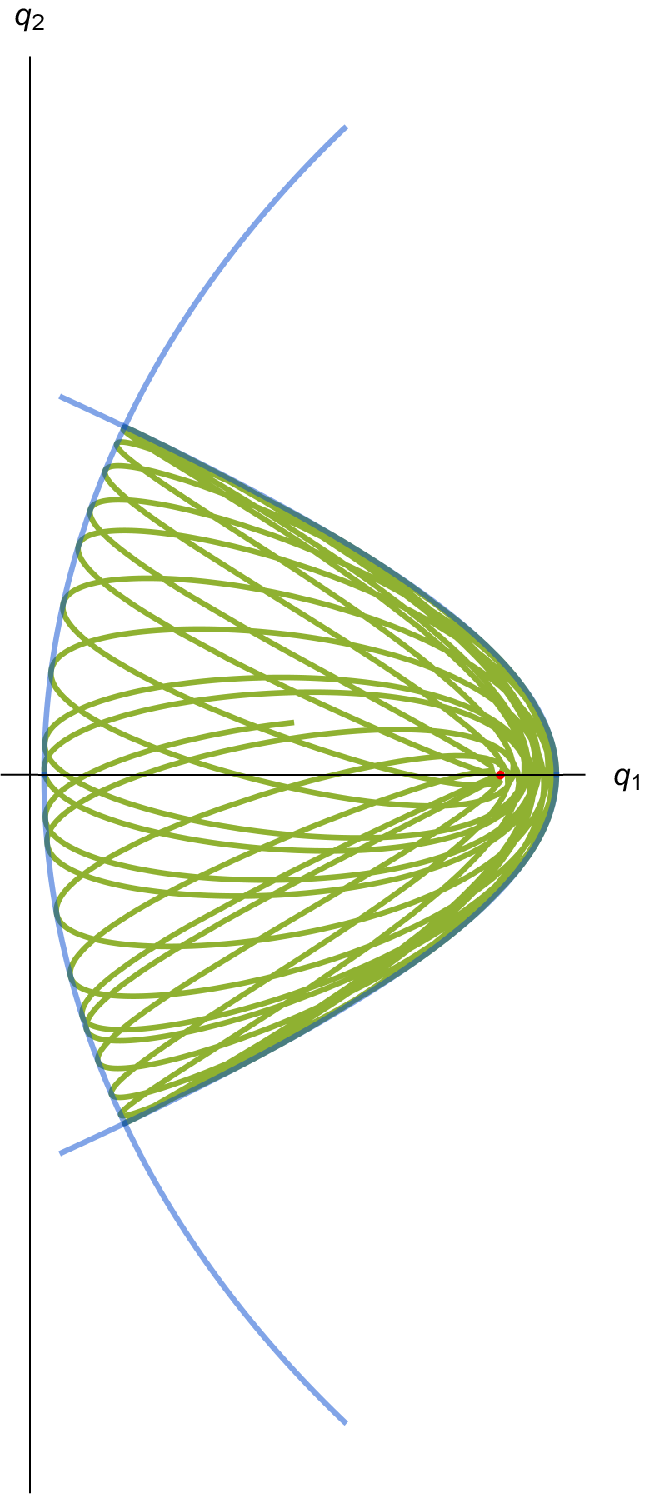} \   \includegraphics[height=2.3cm]{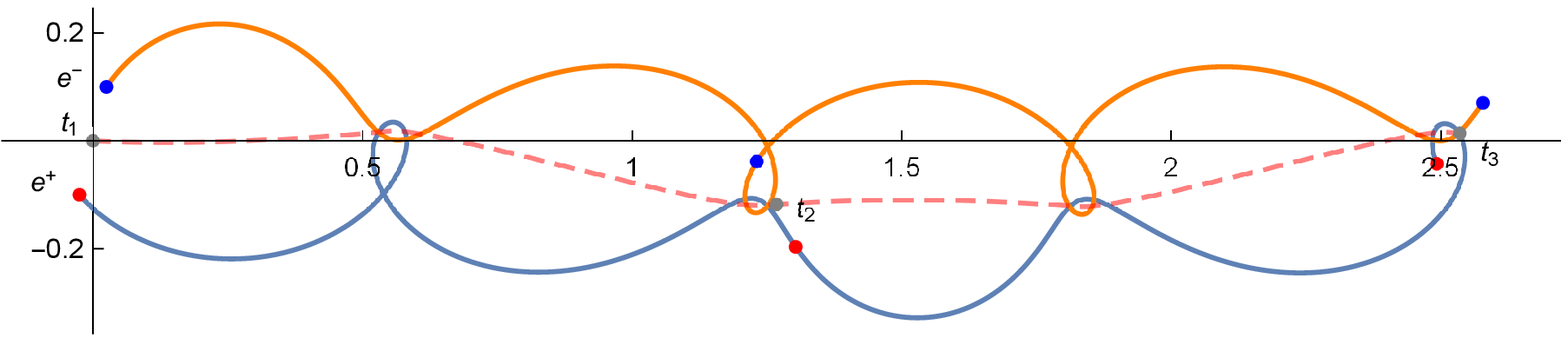}
\caption{$t_{s3}$ orbit with: $\alpha_a=\frac{1}{3}$, $h_a=0.3$, $\lambda_a=0$, $(x_0,y_0)=(0,1)$, $q_1(0)=-1.2$, $q_2(0)=0.05$. Left) Relative particle orbit for $t\in[0,20]$. Right) Electron(orange)-positron(blue) orbits and center of mass trajectory (red-dashed) with $t\in[0,4]$.}
\label{AppEj3}       
\end{center}
\end{figure}


\begin{figure}[h]
\begin{center}
\includegraphics[height=4.5cm]{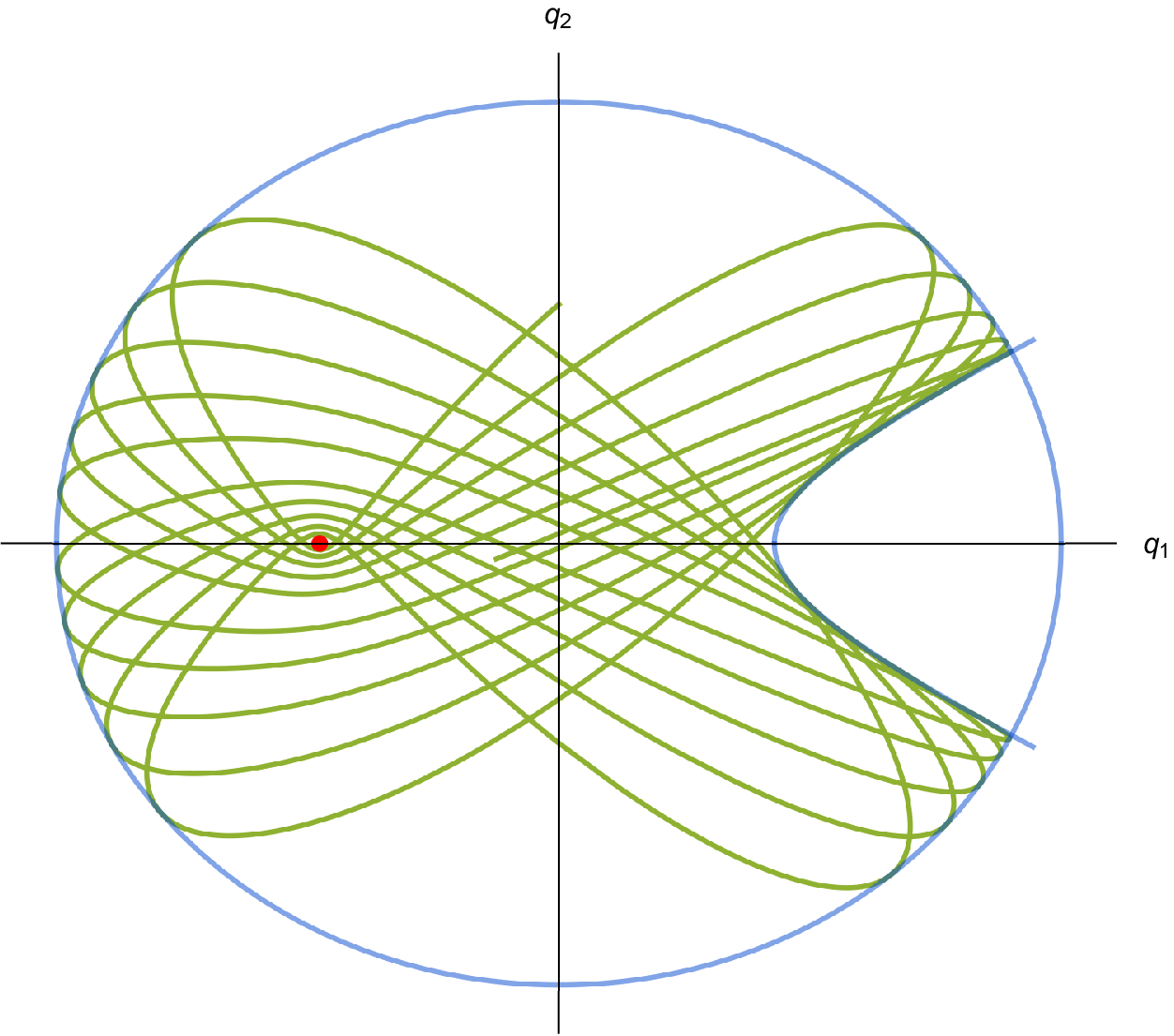} \  \   \includegraphics[height=6.2cm]{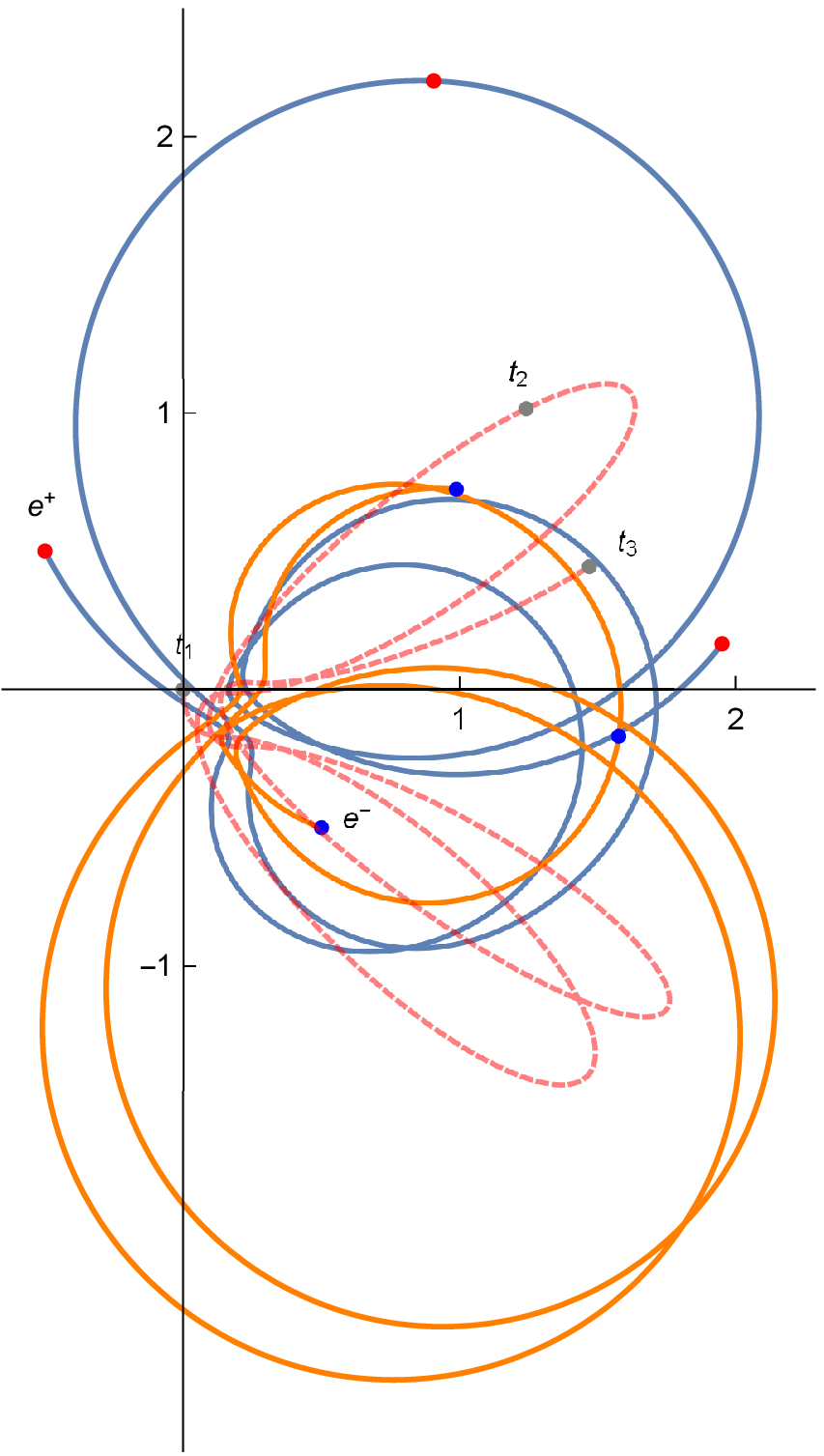}
\caption{$t_{s2}$ orbit with: $\alpha_a=\frac{1}{3}$, $h_a=2$, $\lambda_a=2$, $(x_0,y_0)=(0,1)$, $q_1(0)=0$, $q_2(0)=1$. Left) Relative particle orbit for $t\in[0,60]$. Right) Electron(orange)-positron(blue) orbits and center of mass trajectory (red-dashed) with $t\in[0,20]$.}
\label{AppEj2}
\end{center}
\end{figure}


\begin{figure}[h]
\begin{center}
\includegraphics[height=4.3cm]{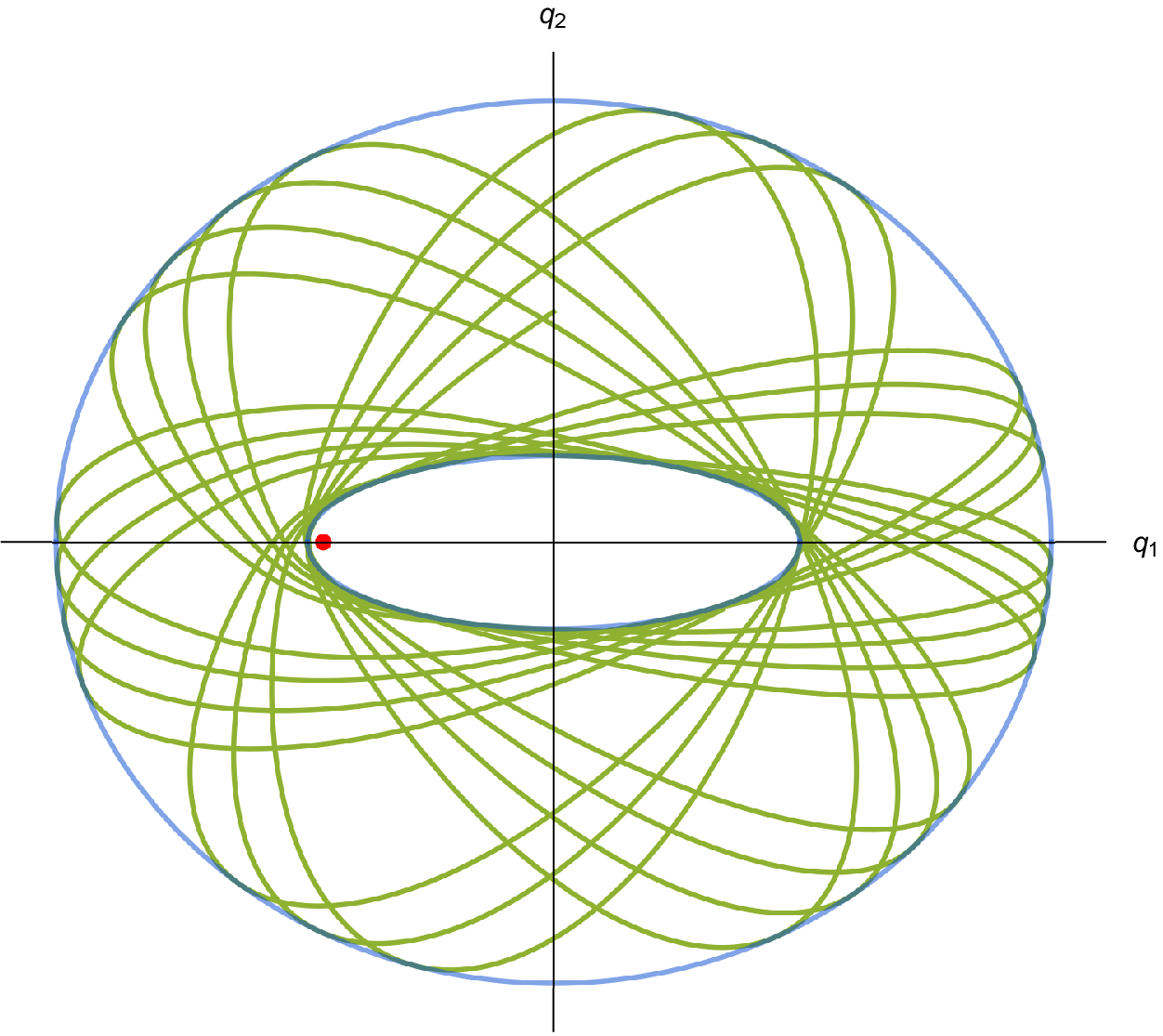} \  \  \includegraphics[height=4.5cm]{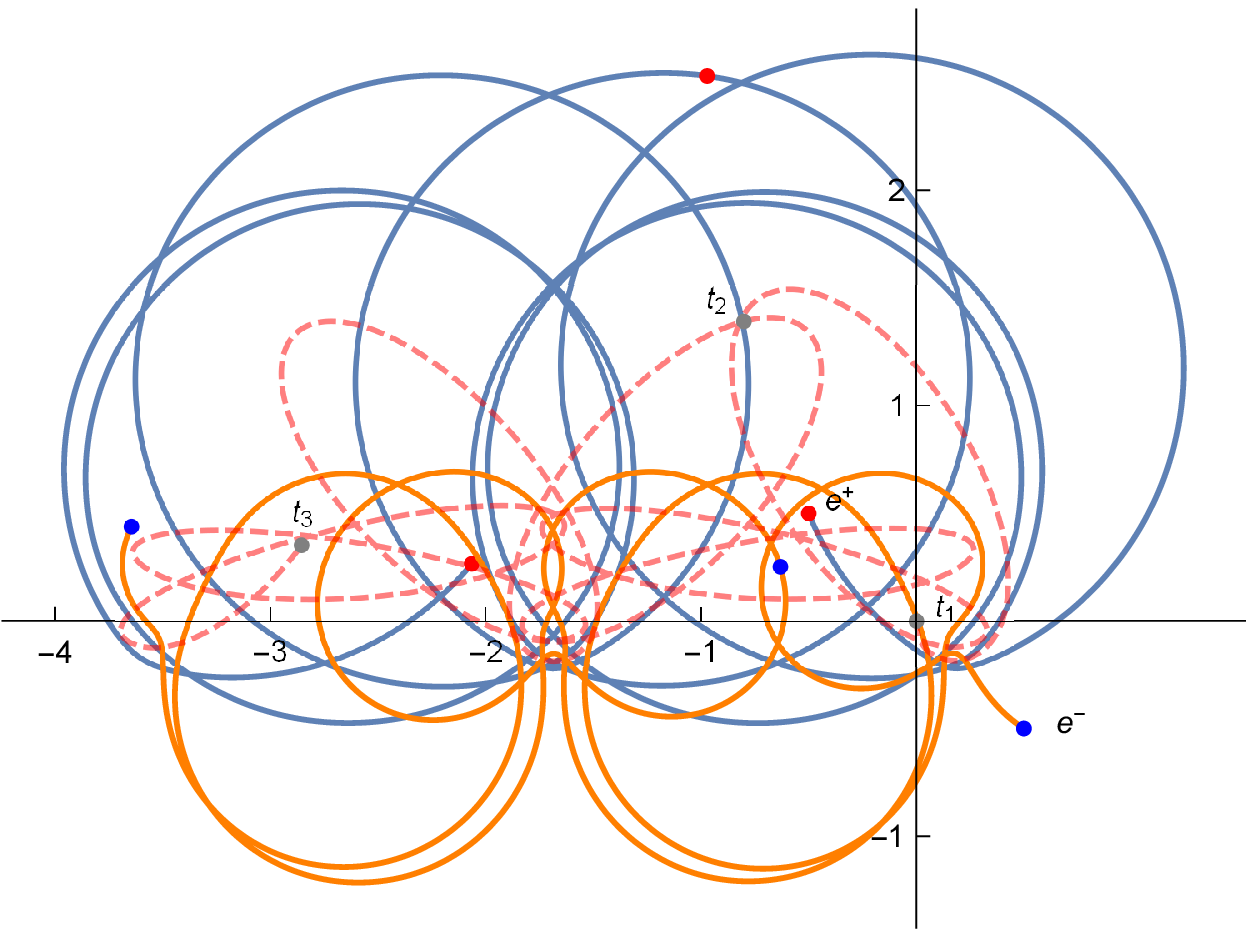}
\caption{$t_{p2}$ orbit with: $\alpha_a=\frac{1}{3}$, $h_a=2.3$, $\lambda_a=2.9$, $(x_0,y_0)=(0,1)$, $q_1(0)=0$, $q_2(0)=1$. Left) Relative particle orbit for $t\in[0,80]$. Right) Electron(orange)-positron(blue) orbits and center of mass trajectory (red-dashed) with $t\in[0,40]$.}
\label{AppEj4}       
\end{center}
\end{figure}


\begin{figure}[h]
\begin{center}
\includegraphics[height=4.cm]{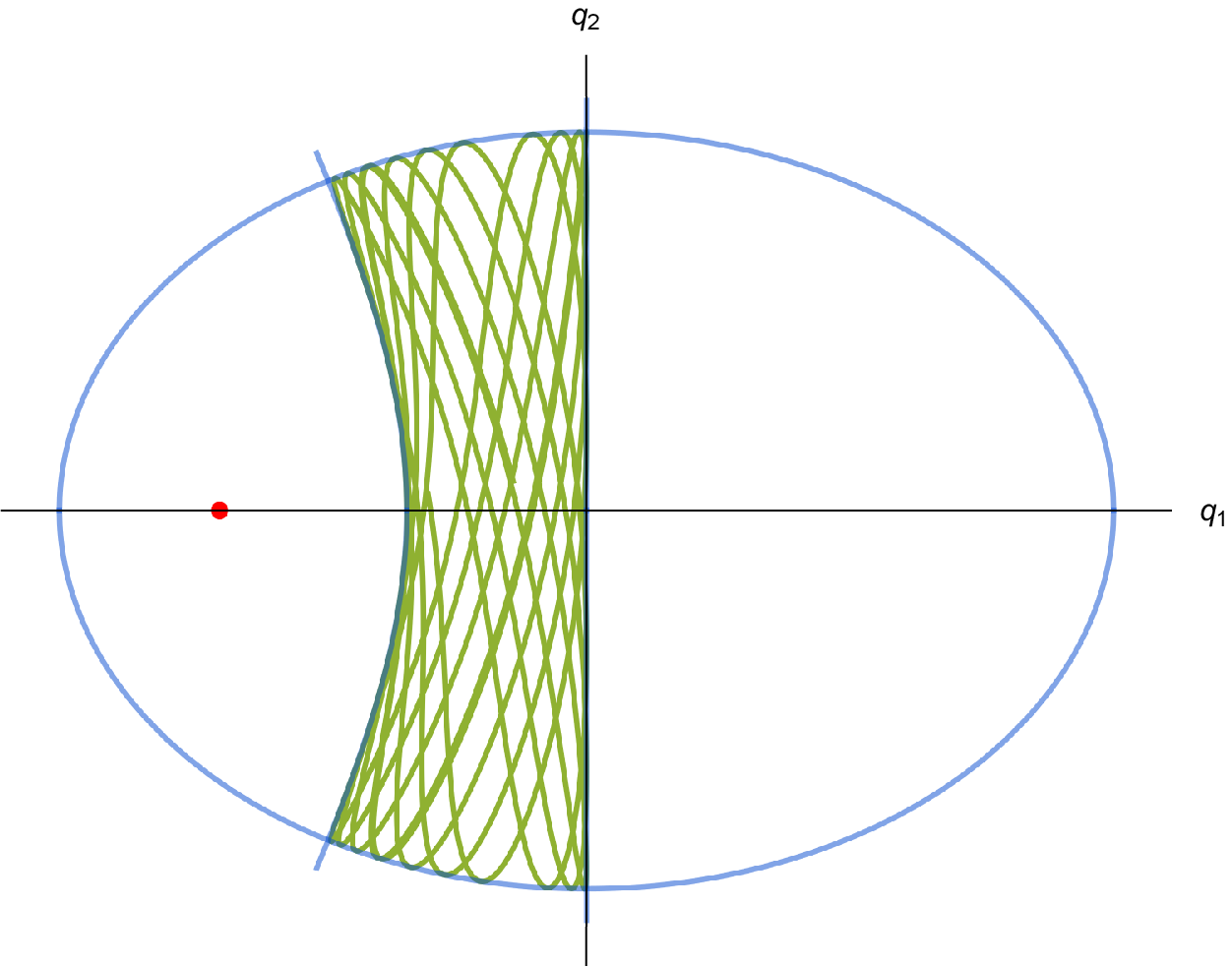} \   \includegraphics[height=3.2cm]{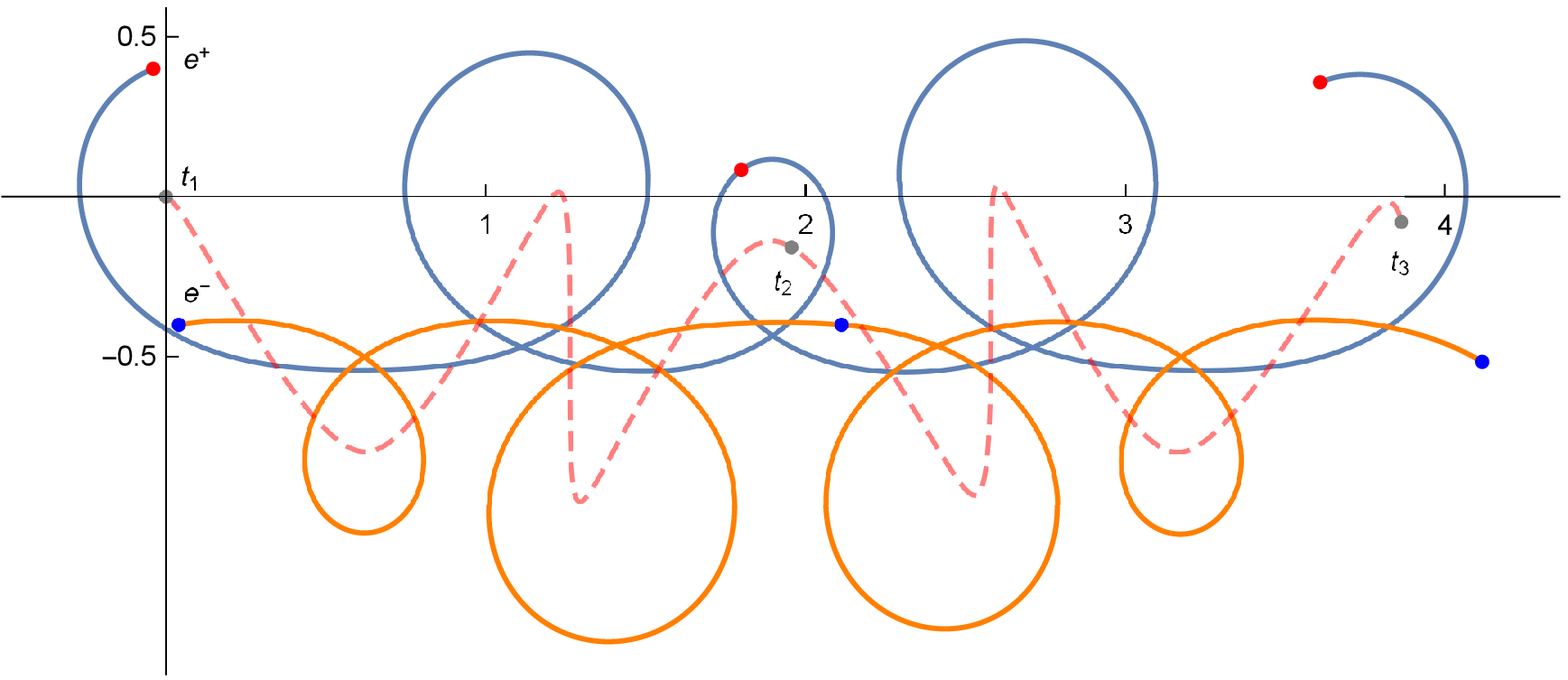}
\caption{$t_{m2}$ orbit with: $\alpha_a=\frac{1}{3}$, $h_a=0.3$, $\lambda_a=0$, $(x_0,y_0)=(0,1)$, $q_1(0)=-0.2$, $q_2(0)=0.08$. Left) Relative particle orbit for $t\in[0,60]$. Right) Electron(orange)-positron(blue) orbits and center of mass trajectory (red-dashed) with $t\in[0,22]$.}
\label{AppEj5}       
\end{center}
\end{figure}


\begin{figure}[h]
\begin{center}
\includegraphics[height=3.5cm]{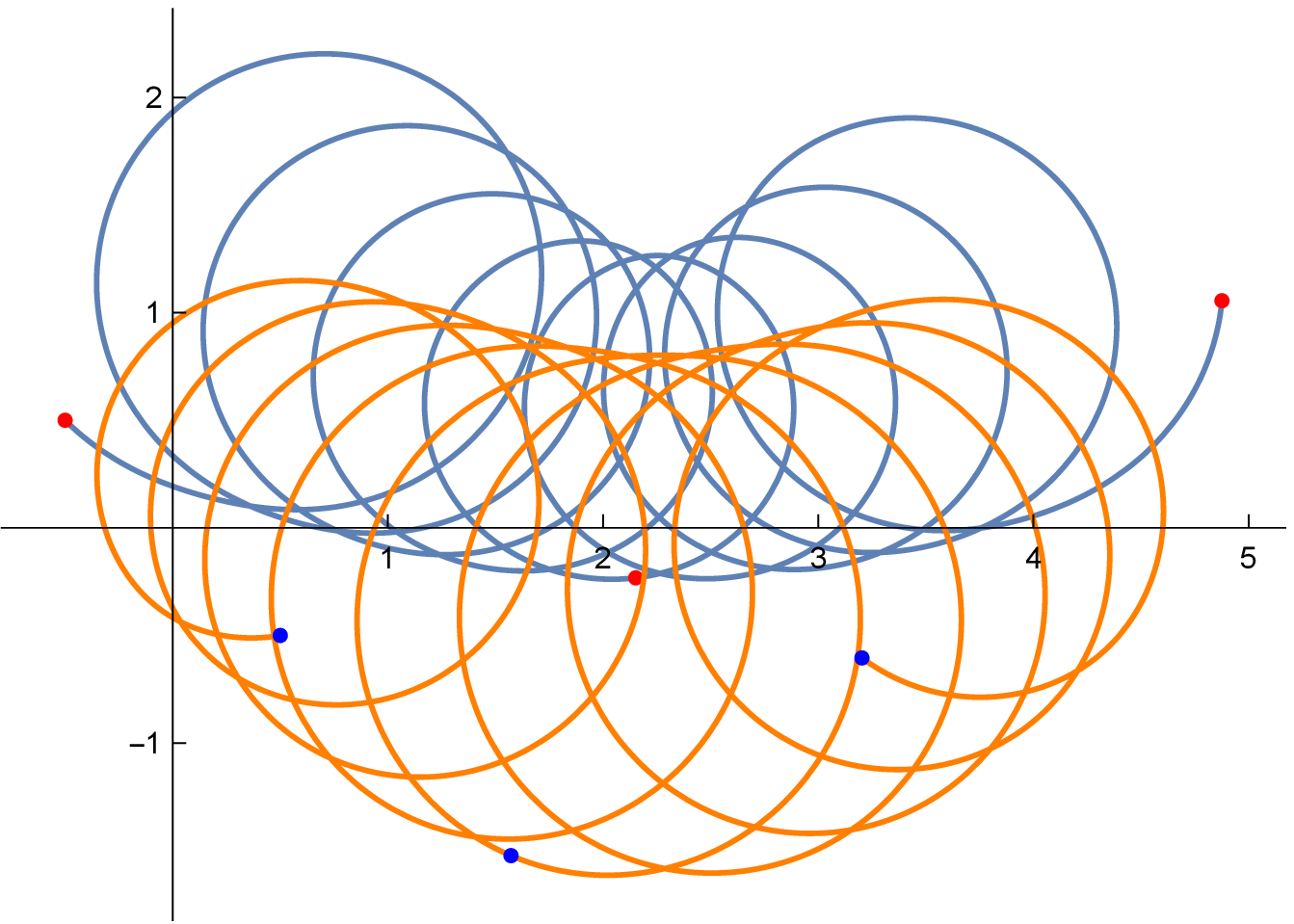} \   \includegraphics[height=2.cm]{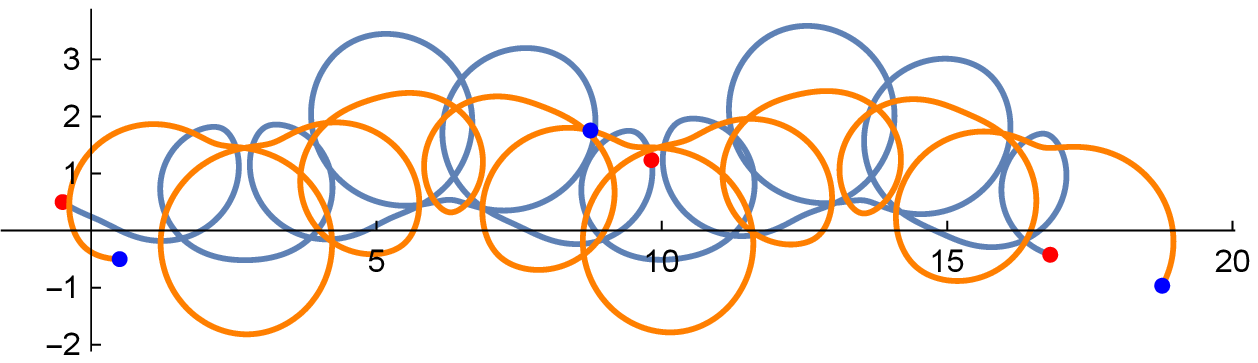}
\caption{Two $t_{m1}$ deformed oscillatory orbits corresponding to different values of $\alpha_a$: Left) $\alpha_a=\frac{1}{3}$, $h_a=2$, $\lambda_a=0.5$, $(x_0,y_0)=(0,1)$, $q_1(0)=0$, $q_2(0)=1$, $t\in[0,50]$. Right) $\alpha_a=2$, $h_a=4$, $\lambda_a=1$, $(x_0,y_0)=(0,1)$, $q_1(0)=0$, $q_2(0)=1$, $t\in[0,50]$.}
\label{AppEj6}       
\end{center}
\end{figure}


\begin{figure}[h]
\begin{center}
\includegraphics[height=3cm]{FigureA14Left} \  \qquad  \quad  \  \includegraphics[height=3.8cm]{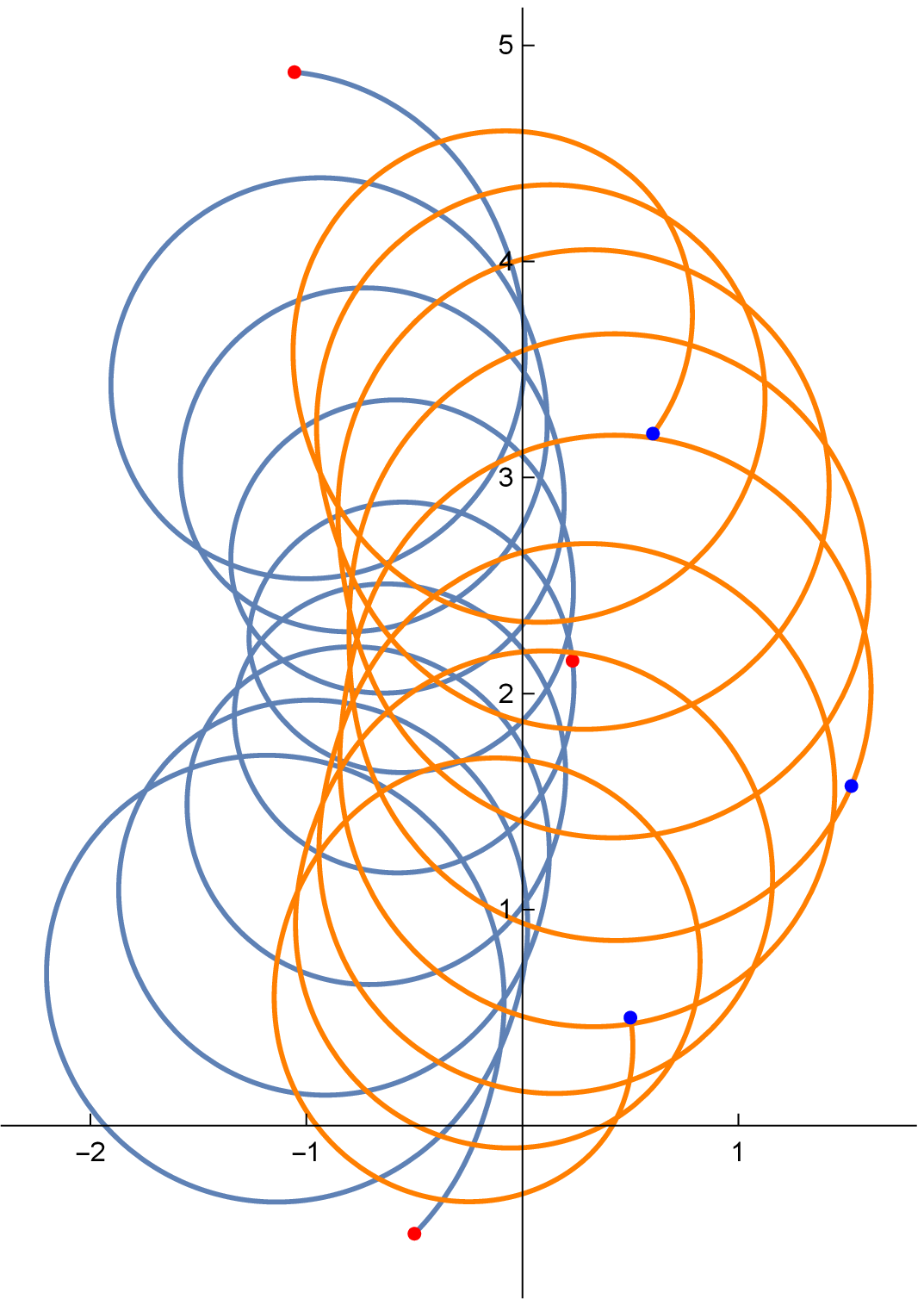} \\
\includegraphics[height=3cm]{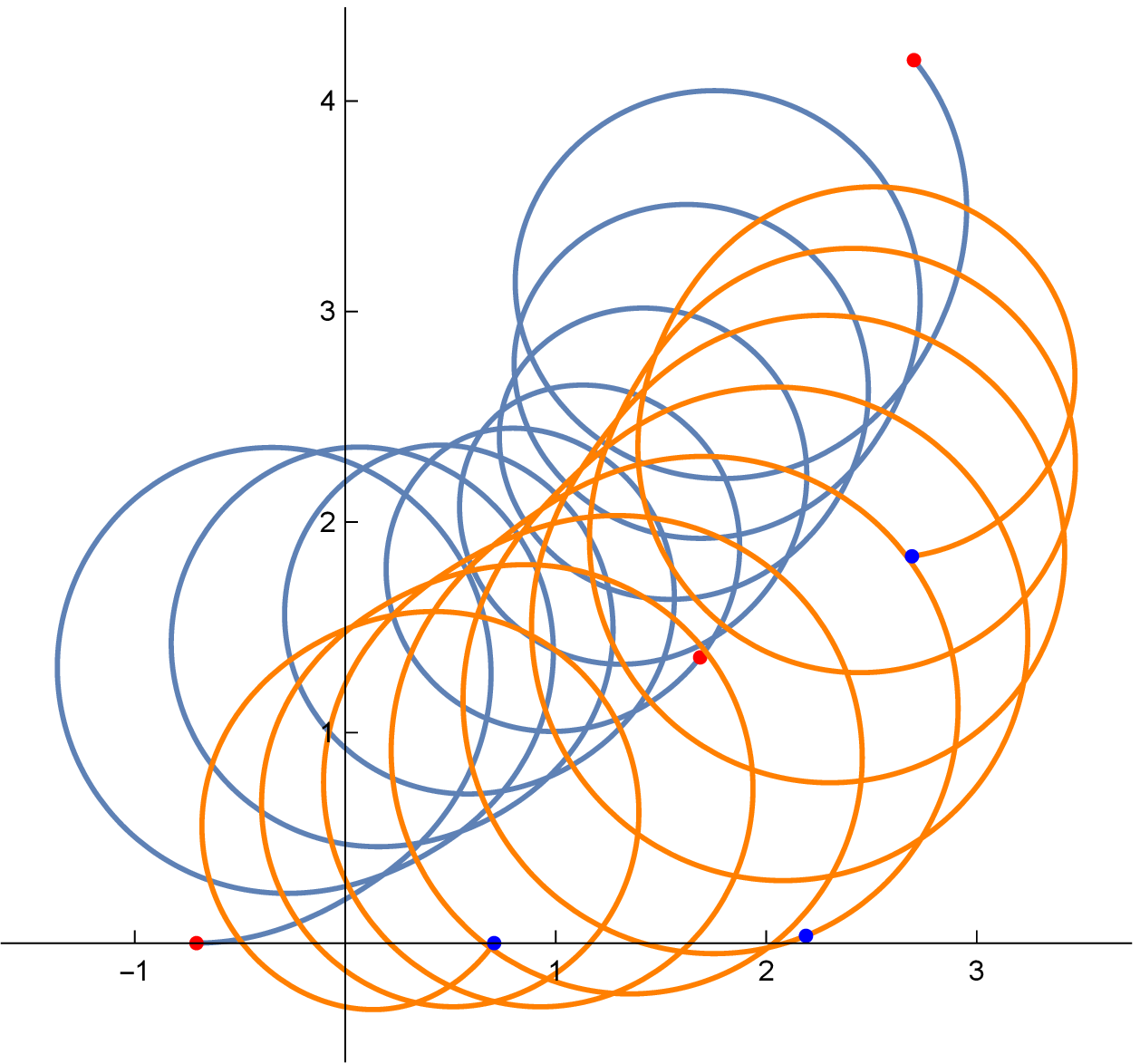} \   \qquad  \quad  \   \includegraphics[height=3cm]{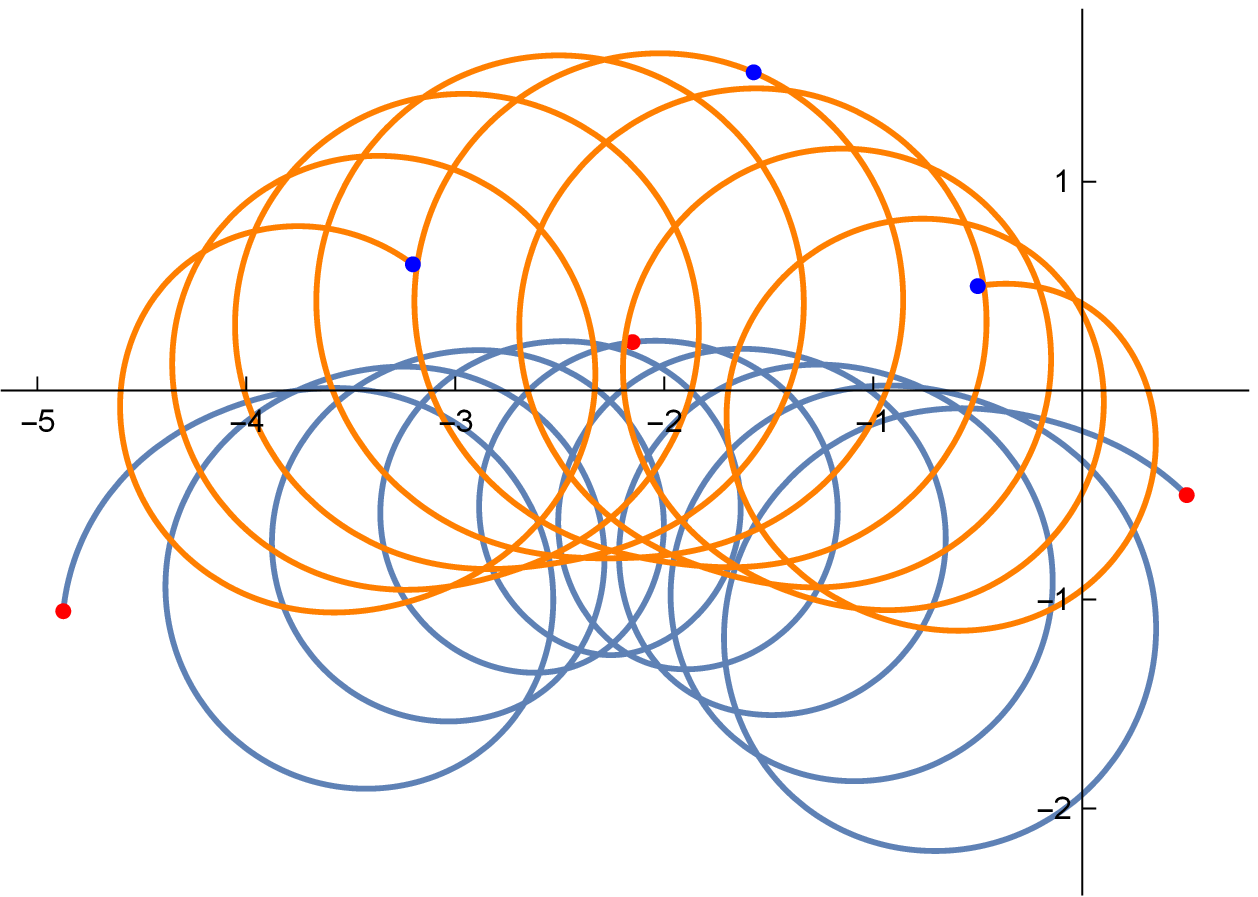}
\caption{Four $t_{m1}$ deformed oscillatory orbits with $\alpha_a=\frac{1}{3}$, $h_a=2$, $\lambda_a=0.5$, $q_1(0)=0$, $q_2(0)=1$, $t\in[0,50]$, and different values of $(x_0,y_0)$: Up-Left) $(x_0,y_0)=(0,1)$, Up-Right) $(x_0,y_0)=(1,0)$, Down-Left) $(x_0,y_0)=\left( \frac{1}{\sqrt{2}},\frac{1}{\sqrt{2}}\right)$, Down-Right) $(x_0,y_0)=(0,-1)$.}
\label{AppEj7}       
\end{center}
\end{figure}


\begin{figure}[h]
\begin{center}
\includegraphics[height=3cm]{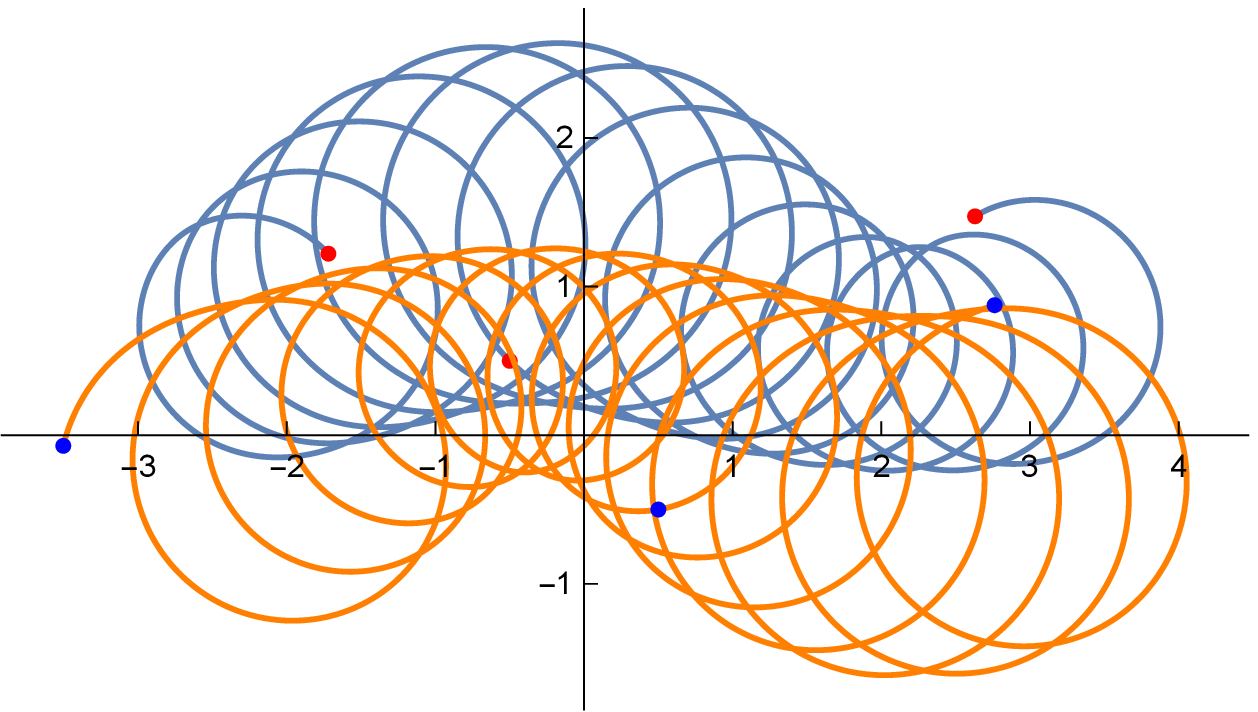}    \  \qquad    \  \includegraphics[height=3cm]{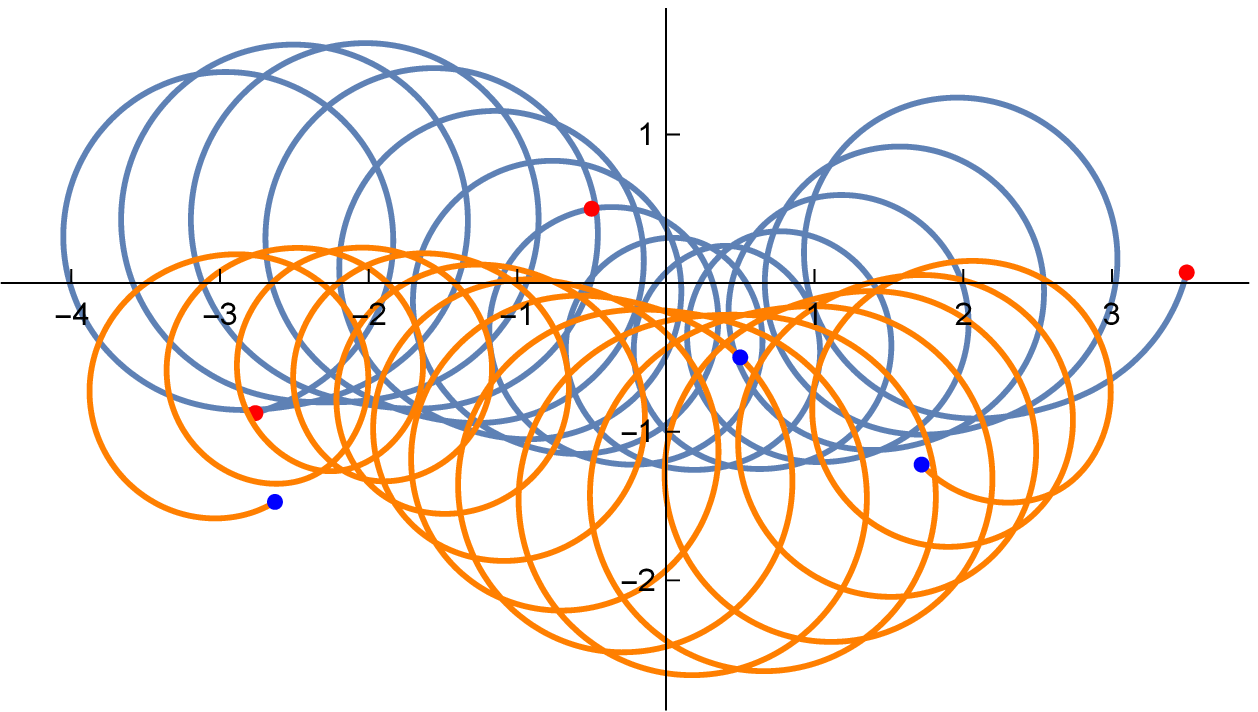} \\
\includegraphics[height=3cm]{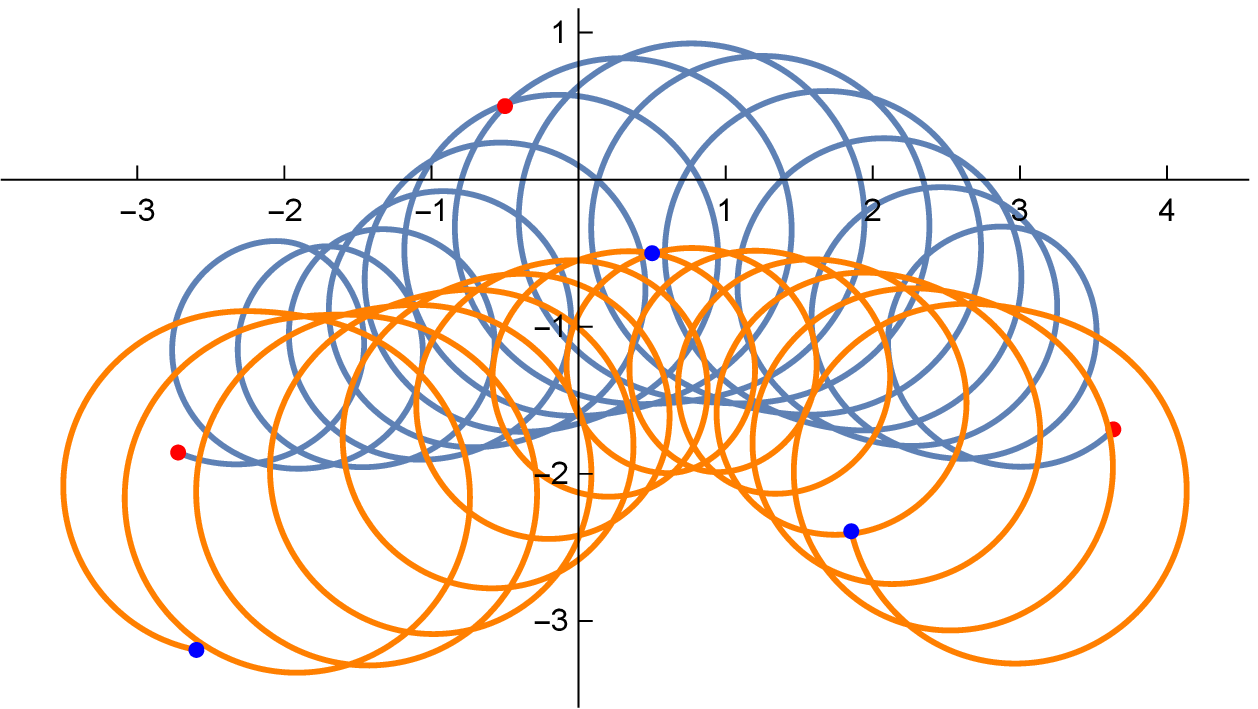}   \  \qquad    \    \includegraphics[height=3cm]{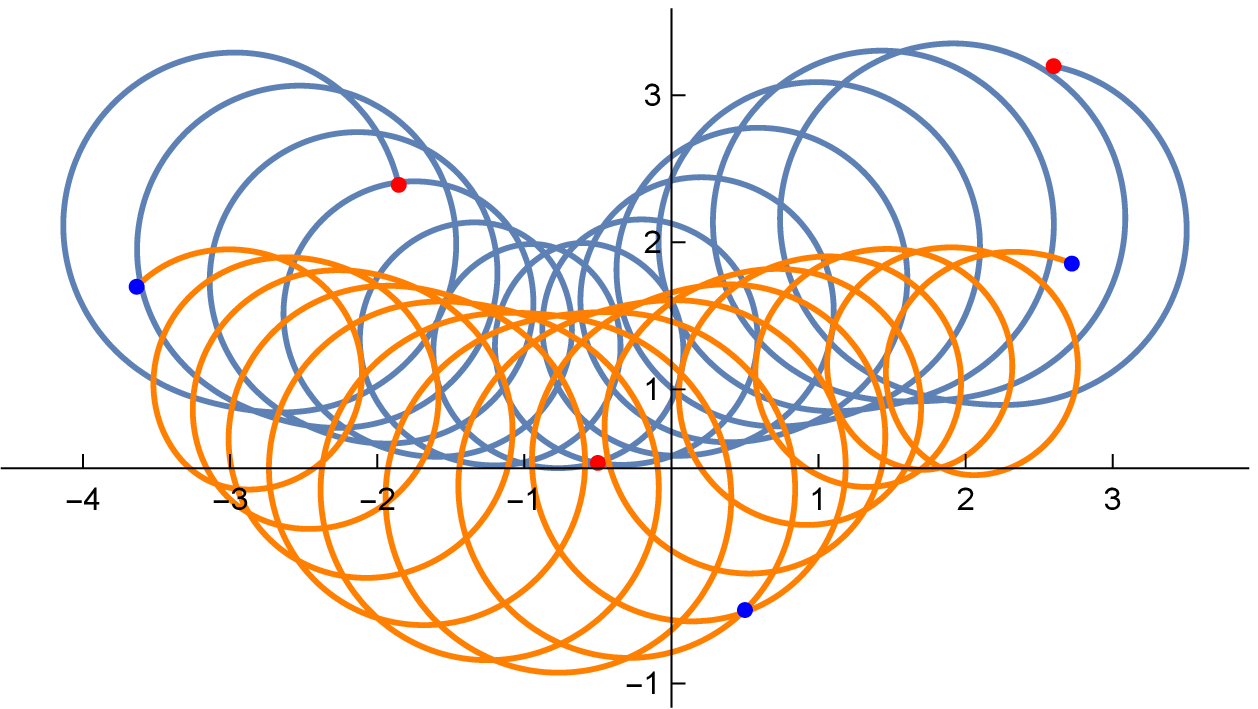}
\caption{Four $t_{m1}$ deformed oscillatory orbits with the same parameters: $\alpha_a=\frac{1}{3}$, $h_a=2$, $\lambda_a=0.5$, $q_1(0)=0$, $q_2(0)=1$, $(x_0,y_0)=(0,1)$, $t\in[-40,40]$, but corresponding to the four different possible choices of $\dot{q}_1(0)$ and $\dot{q}_2(0)$ determined from the same pair $(h_a,\lambda_a)$.}
\label{AppEj8}       
\end{center}
\end{figure}


\end{document}